\documentclass[twocolumn,times,tighten]{aastex631}

\usepackage{xspace}
\usepackage{hyperref}
\usepackage{CJK}

\newcommand{\msun}{M$_{\odot}$}
\newcommand{\rsun}{R$_{\odot}$}
\newcommand{\teff}{${T}_{\mathrm{eff}}$}

\newcommand{\eqwk}{$\mathrm{W_{CaK}}$}
\newcommand{\eqwfe}{$\mathrm{W_{FeII-5169}}$}
\newcommand{\eqwna}{$\mathrm{W_{D_1+D_2}}$}

\newcommand{\obj}{ZTF\,J1944+4557\xspace}
\newcommand{\code}[1]{\texttt{#1}}

\shorttitle{5-hr Transits at ZTF\,J1944+4557}
\shortauthors{Guidry \& Vanderbosch et al.}

\received{June 10, 2025}
\revised{August 22, 2025}
\accepted{August 22, 2025}

\submitjournal{ApJ}

\graphicspath{{./}{figures/}}
\begin{document}

\title{Transiting Planetary Debris near the Roche Limit of a White Dwarf on a 4.97\,hr Orbit -- and its Vanishing}

\author[0000-0001-9632-7347]{Joseph A. Guidry}\altaffiliation{NSF Graduate Research Fellow}
\affiliation{Department of Astronomy \& Institute for Astrophysical Research, Boston University, Boston, MA 02215, USA}

\author[0000-0002-0853-3464]{Zachary P. Vanderbosch}
\affil{Department of Astronomy, California Institute of Technology, 1216 E. California Blvd, Pasadena, CA, 91125, USA}

\collaboration{10}{The lead authors above contributed equally to the manuscript and are listed in alphabetical order.}

\author[0000-0001-5941-2286]{J. J. Hermes}
\affiliation{Department of Astronomy \& Institute for Astrophysical Research, Boston University, Boston, MA 02215, USA}

\author[0000-0001-8014-6162]{Dimitri Veras}
\affiliation{Centre for Exoplanets and Habitability, University of Warwick, Coventry CV4 7AL, UK}
\affiliation{Centre for Space Domain Awareness, University of Warwick, Coventry CV4 7AL, UK}
\affiliation{Department of Physics, University of Warwick, Coventry CV4 7AL, UK}

\author[0000-0003-0089-2080]{Mark A. Hollands}
\affiliation{Department of Physics, University of Warwick, Coventry CV4 7AL, UK}

\author[0000-0003-2071-2956]{Soumyadeep Bhattacharjee}
\affiliation{Department of Astronomy, California Institute of Technology, 1216 E. California Blvd, Pasadena, CA, 91125, USA}

\author[0000-0002-4770-5388]{Ilaria Caiazzo}
\affiliation{Institute of Science and Technology Austria, Am Campus 1, 3400 Klosterneuburg, Austria}

\author[0000-0002-6871-1752]{Kareem El-Badry}
\affiliation{Department of Astronomy, California Institute of Technology, 1216 E. California Blvd, Pasadena, CA, 91125, USA}

\author[0000-0001-5745-3535]{Malia L. Kao}
\affiliation{Department of Astronomy, University of Texas at Austin, 2515 Speedway, Austin, TX 78712, USA}

\author[0009-0002-6065-3292]{Lou~Baya Ould~Rouis}
\affiliation{Department of Astronomy \& Institute for Astrophysical Research, Boston University, Boston, MA 02215, USA}

\author[0000-0003-4189-9668]{Antonio C. Rodriguez}
\affiliation{Department of Astronomy, California Institute of Technology, 1216 E. California Blvd, Pasadena, CA, 91125, USA}

\author[0000-0002-2626-2872]{Jan van~Roestel}
\affiliation{Anton Pannekoek Institute for Astronomy, University of Amsterdam, 1090 GE Amsterdam, The Netherlands}

\correspondingauthor{Joseph Guidry, Zachary Vanderbosch}
\email{jaguidry@bu.edu; zvanderbosch@gmail.com}

\begin{abstract}

We present the discovery of deep, irregular, periodic transits towards the white dwarf ZTF\,J1944$+$4557 using follow-up time-series photometry and spectroscopy from Palomar, Keck, McDonald, Perkins, and Lowell observatories. We find a predominant period of 4.9704\,hr, consistent with an orbit near the Roche limit of the white dwarf, with individual dips over 30\% deep and lasting between 15 and 40 minutes. Similar to the first known white dwarf with transiting debris, WD\,1145$+$017, the transit events are well-defined with prominent out-of-transit phases where the white dwarf appears unobscured. Spectroscopy concurrent with transit photometry reveals the average Ca\,K equivalent width remains constant in and out of transit. The broadening observed in several absorption features cannot be reproduced by synthetic photospheric models, suggesting the presence of circumstellar gas. Simultaneous $g+r$- and $g+i$-band light curves from the CHIMERA instrument reveal no color dependence to the transit depths, requiring transiting dust grains to have sizes $s \gtrsim0.2\,\mu$m. The transit morphologies appear to be constantly changing at a rate faster than the orbital period. Overall transit activity varies in the system, with transit features completely disappearing during the seven months between our 2023 and 2024 observing seasons and then reappearing in 2025~March, still repeating at 4.9704\,hr. Our observations of the complete cessation and resumption of transit activity provide a novel laboratory for constraining the evolution of disrupted debris and processes like disk exhaustion and replenishment timescales at white dwarfs.

\end{abstract}


\section{Introduction} \label{sec:intro}

White dwarf stars mark the end stage of stellar evolution for essentially all known planet-hosting stars, and are frequent sites of active destruction of exoplanetary objects. More than 40\% of white dwarf atmospheres are enriched with metals accreted from tidally disrupted planetesimals \citep{Zuckerman2003,Zuckerman2010,Koester2014,Wilson2019,OuldRouis2024}. Evidence for this circumstellar shredding is also found as excess infrared emission from warm dust \citep[e.g.,][]{Rocchetto2015} and broad, double-peaked optical emission from metallic gas \citep[e.g.,][]{Manser2020}, all orbiting near to or within the tidal disruption radius of the star. These observables are modeled well by a reservoir of planets that survive post-main sequence evolution and perturb smaller, asteroid-sized bodies towards the host white dwarf \citep{Debes2002}, close enough to be tidally disrupted and fill out a disk that can be accreted \citep{Jura2003}. We refer readers to \citet{Farihi2016} and \citet{Veras2024} for thorough reviews of this picture.

Large time-domain surveys in the 2010s finally captured observations of transits by tidally disrupted extrasolar planetesimals. The extended Kepler mission, K2, first observed transits at the white dwarf WD\,1145+017 \citep{Vanderburg2015}, where irregular dimmings of up to 40\% repeat every 4.5 hours. The measurement of the orbital period of the transiting debris around WD\,1145+017 is as significant as the confirmation of an infrared excess \citep{Vanderburg2015,Xu2018a} and the detection of circumstellar gas \citep{Xu2016} in the system; the 4.5-hour period corresponds to a circular orbit near the Roche limit for a Solar System asteroid-like body. WD\,1145+017 corroborated the working tidal disruption model as an explanation for signposts of planetesimal destruction at white dwarfs \citep[e.g.,][]{vanLieshout2018}.

Transiting debris at white dwarfs has since been observed in ever-increasing numbers. Three additional white dwarfs have been discovered to be repeatedly occulted by debris at longer periods that imply eccentric orbits extending beyond the Roche limit: ZTF\,J0328$-$1219 shows recurring transits at 9.937\,hr and 11.2\,hr \citep{Vanderbosch2021}, WD\,1054$-$226 at 25.02\,hr \citep{Farihi2022}, and ZTF\,J0139+5245 at roughly 107\,d \citep{Vanderbosch2020}. TESS Sector~75 photometry of SBSS\,1232+563 captured a coherent 14.842\,hr signal subsequent to the deep, eight months-long transit event in 2023, which might reflect the period of the bulk orbit of the transiting debris \citep{Hermes2025}. These detected orbital periods are consistent with predicted orbital period distributions for gravitationally scattered and tidally disrupted asteroids around white dwarfs \citep{Li2025a}. \citet{Guidry2021} discovered three additional transiting systems by searching for white dwarfs with anomalously large photometric scatter in Gaia and Zwicky Transient Facility (ZTF) photometry, although orbital periods towards these three objects have yet to be measured. \citet{Bhattacharjee2025} expanded this technique to discover six more new transiting systems, again without measuring periodicities. We provide a summary of the properties of the white dwarfs that show periodic transits from planetary debris in Section~\ref{sec:conclusion}.

Transits appear to demonstrate several physical effects driven by the dynamics of tidally disrupted debris at white dwarfs. Collisional cascades are expected to grind down planetesimals into the dust viewed as an infrared excess \citep{KenyonBromley2017a,KenyonBromley2017b,Brouwers2022}. Near-infrared variability corroborates this picture \citep{Swan2020, Swan2021, Guidry2024, Noor2025}, as do transits. WD\,1145+017, ZTF\,J0328$-$1219, and WD\,1054$-$226 all show additional periodicities independent of the dominant period in their photometry. Collisions likely explain observed drifting fragments that are perturbed onto independent orbits, as well as observed variability in transit morphologies \citep{Gaensicke2016,Rappaport2016,Gary2017,Vanderbosch2021,Farihi2022}.

Transit activity levels often appear to be variable or transient. SBSS\,1232+563 shows sporadic, deep transits that repeat every 6-10 years across 25 years of monitoring, while otherwise only showing flux excursions at the percent level during quiescent, out-of-deep-transit phases  \citep{Hermes2025}. WD\,1145+017 and ZTF\,J0328$-$1219 have shown varying phases of transit activity and depths \citep{Gary2017,Aungwerojwit2024}. WD\,1145+017 has recently stopped showing deep transits repeating at 4.5 hours \citep{Aungwerojwit2024}. The range in transit recurrence timescales is also evidence for the expected circularization of disrupted material \citep{Brouwers2022,Li2025a}: tidal interactions should draw in initially disrupted material from extremely eccentric ($e \gtrsim 0.9$), au-scale orbits with periods of at least hundreds of days onto closer-in, less eccentric orbits with semi-major axes near the star's Roche limit, typically about 1\,\rsun for white dwarfs.

We present in this manuscript detailed observations of one of the six transiting white dwarfs discovered by \citet{Bhattacharjee2025}, ZTF\,J194431.92+455753.13 (Gaia DR3 2080201713998110336; hereafter \obj), which show evidence for the dynamical interactions of disrupted planetesimals. After confirming \obj\ as a transiting debris system due to excess Gaia and ZTF photometric scatter, negative skew in the ZTF light curve, and atmospheric metal pollution, \citet{Bhattacharjee2025} obtained follow-up high-speed time-series photometry that showed a single deep eclipse lasting about $\approx$\,20\,minutes. We add extensive follow-up photometry covering a baseline of over 3.3 years that show these transits repeat roughly every 4.97\,hr, marking the second case of repeating transits near the Roche limit of a white dwarf. Our follow-up further shows a vanishing of the repeating transits beginning as late as 2024~July and lasting through at least 2024~October, and a re-appearance of transits in 2025~March. 

We detail our follow-up time-series photometry campaign and our spectroscopic monitoring of \obj\ in Section~\ref{sec:obs}, provide our analysis of the periodicity of the transits in Section~\ref{sec:period}, assess the metal enrichment of our spectroscopy of \obj\ in Section~\ref{sec:pollution}, review the properties of the transits at \obj\ in Section~\ref{sec:discussion}, and recapitulate our main conclusions in Section~\ref{sec:conclusion}.

\section{Observations} \label{sec:obs}

\subsection{High-Speed, Time-Series Photometry}\label{sec:timeseries}

\begin{figure*}[t!]
    \epsscale{1.17}
    \plotone{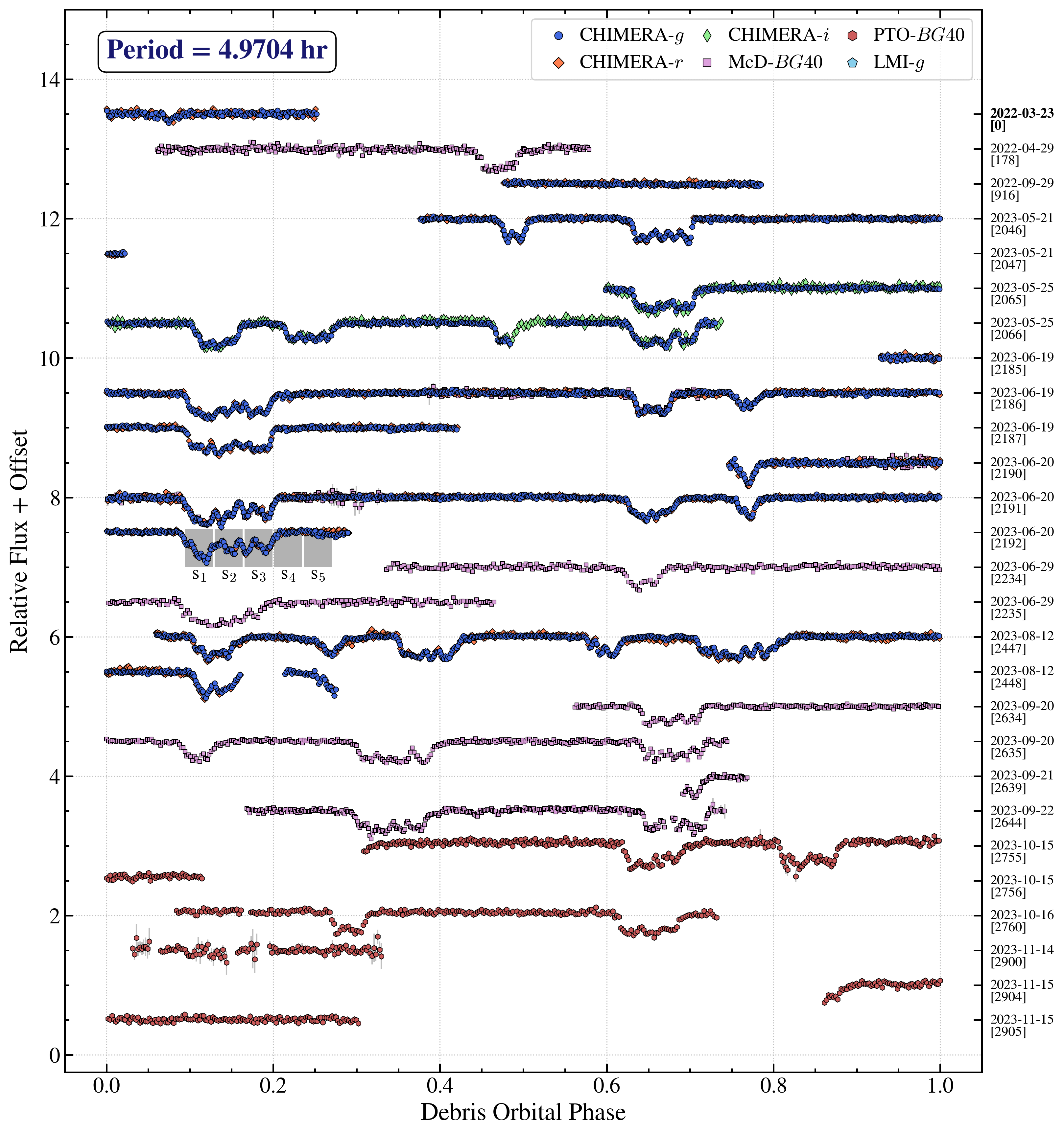}
    \caption{Follow-up time-series photometry from 2022~March through 2023~November for ZTF\,J1944+4557 from Palomar ($g$-, $r$-, and $i$-bands), McDonald ($BG40$-band), Perkins ($BG40$-band), and Lowell ($g$-band) observatories, folded on a period of 4.9704\,hr. Light curves are shown in chronological order from top to bottom, with each row representing a single orbital cycle. The UTC observation dates for each row are shown at right along with orbital cycle numbers in brackets, with phase\,${=}\,0$ during cycle$\,{=}\,0$ corresponding to $\mathrm{BJD_{TDB}}\,{=}\,2459661.9693637$. The shaded regions overlapping the 2023 June 20 (Cycle 2192) light curve represent the times during which we acquired five consecutive spectra with Keck LRIS (see Figure~\ref{fig:EW_transit}). The dual-channel CHIMERA photometry stringently constrain the transits as colorless (see Figure~\ref{fig:color_dependence}). In-transit variations evolve from one cycle to the next, yet the ingresses and egresses from consecutive nights align on the 4.9704\,hr orbital period. The location in phase of the transits appears dynamic on a timescale of $\mathcal{O}(10)$~cycles. As of 2024~May~14 (Cycle 3780) the transits appear to have vanished until re-appearing in 2025~March (see Figure~\ref{fig:phasedLCs_24}). Our LMI photometry are shown only in Figure~\ref{fig:phasedLCs_24} (see Appendix~\ref{sec:ts_phot_appendix}). \label{fig:phasedLCs_22-23}}
\end{figure*}

\begin{figure*}[t!]
    \epsscale{1.17}
    \plotone{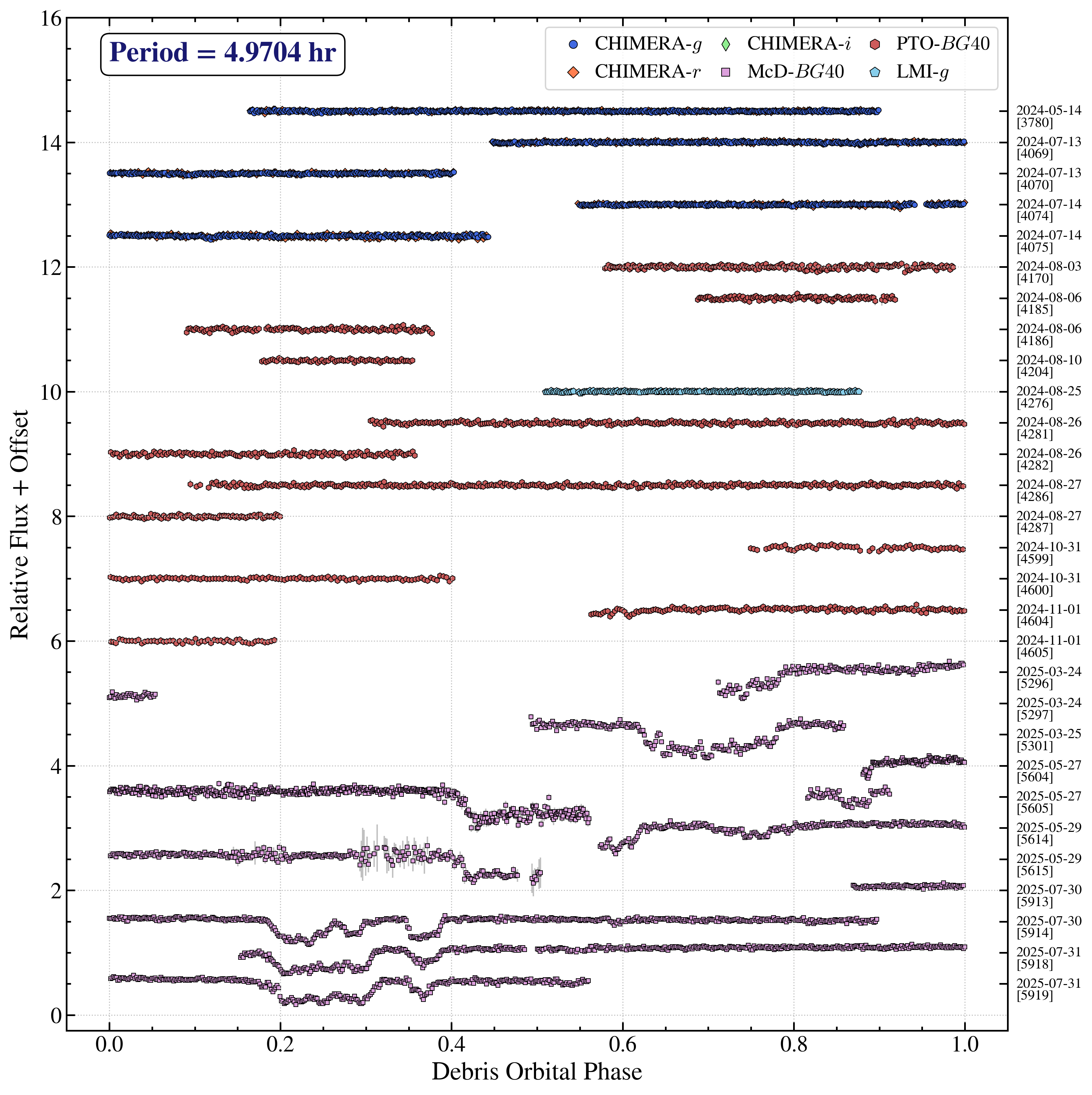}
    \caption{Same as Figure~\ref{fig:phasedLCs_22-23} but showing follow-up time-series photometry from 2024~May--November and 2025~March--July. Aside from a possible small dip at the beginning of 2024~November~1 observations, no transit features are observed in 2024. Transits were again observed in 2025~March, with the two observed egresses phasing together on $P=4.9704$\,hr. We similarly do not resolve entire transits in our 2025~May observations due to inclement weather, but do capture two egresses and two ingresses that still align at the orbital period. Our uninterrupted 2025~July epochs fully resolve transit recurrence. \label{fig:phasedLCs_24}}
\end{figure*}

We obtained 123.2~hours of follow-up, high-speed, time-series photometry from 31 nights spread across a baseline extending over about 3.3 years using facilities at Palomar Observatory, McDonald Observatory, the Perkins Telescope Observatory, and the Lowell Discovery Telescope. This does not count overlapping concurrent runs from multiple facilities. These photometry includes the three light curves from \citet{Bhattacharjee2025} that confirmed \obj\ as a new transiting debris system. Our photometry, regardless of their instrument, are reduced consistently using the routines provided by the {\tt hipercam}\footnote{\url{https://github.com/HiPERCAM/hipercam}} photometry reduction pipeline \citep{Dhillon2021}, which we describe below in detail. We waterfall the light curves from 2022-2023 observing runs in Figure~\ref{fig:phasedLCs_22-23} and 2024-2025 observing runs in Figure~\ref{fig:phasedLCs_24}, phased on the best period of 4.9704\,hr (see Section~\ref{sec:period}). A log of our observations is presented in Appendix~\ref{sec:ts_phot_appendix}.

For each light curve obtained we used {\tt hipercam} to perform PSF photometry. We selected {\tt variable} apertures and {\tt variable} PSF profiles with the {\tt optimal} extraction mode. Circular annuli were drawn 20$-$30\,pixels from the centroids of our target and selected comparison stars. We modified certain parameters, such as {\tt search\_half\_width} depending on the sky conditions to optimize the photometry for a given night. We used the \code{phot2lc} Python package \citep{phot2lc} to create the final divided light curves, clipping 5$\sigma$ outliers and removing long-term trends by fitting a low-order (linear or quadratic) polynomial to the light curve, unless this normalization contorted the flat out-of-transit continua. We applied barycentric corrections to the GPS timestamps of our images using {\tt Astropy} \citep{astropy:2013,astropy:2018,astropy:2022}.

{\bf Palomar Observatory (P200)} --- We carried out time series photometry at Palomar Observatory on ten separate nights totaling 45.86\,hr using the Caltech High Speed Multi-color Camera (CHIMERA, \citealt{Harding2016}) attached at prime focus of the 200-in Hale Telescope. We used the $g$- and $r$-band filters on the blue and red arms of the instrument, respectively, on all nights except 2023 May 25 when we used the $i$-band filter on the red arm. We used exposure times of 10\,s for all observations and used standard calibration frames taken each night to bias and flat-field correct our images. Read-out times with CHIMERA are near-instantaneous ($\sim\,1$\,ms). We performed PSF photometry using {\tt hipercam} and used \code{phot2lc} to extract the optimal light curve.

{\bf McDonald Observatory (McD)} --- We observed \obj at the 2.1-m Otto Struve Telescope at McDonald Observatory using the Princeton Instruments ProEM frame-transfer CCD mounted at Cassegrain focus. Our 35.62~total hours of observations over 13 nights cover a 1.6\arcmin$\times$1.6\arcmin~field of view at a plate scale of $0.36$\arcsec~per pixel using 4$\times$4~binning. All exposures were collected using the broad, blue-bandpass, red-cutoff \emph{$BG40$} filter to reduce sky noise. ProEM readout times are near-instantaneous ($\approx\,2$\,ms). We reduced the images using bias, dark, and flat-field calibration images using standard Python routines and the {\tt Astropy}-affiliated {\tt ccdproc} \citep{ccdproc_2017} suite of tools. We again perform PSF photometry using {\tt hipercam} and used \code{phot2lc} to extract the optimal light curve.

{\bf Perkins Telescope Observatory (PTO)} --- We observed \obj\ for 33.35 total hours over 11 nights at the 72-in Perkins Telescope Observatory (PTO) using the Perkins Re-Imaging SysteM (PRISM; \citealt{Janes2004}) mounted at Cassegrain focus. Our PRISM images typically cover a 2.5\,square~arcminute field of view at a plate scale of $0.39''$~per pixel after windowing, which reduces the overhead from readout to $\approx\,5$\,s. We exclusively used the \emph{BG40} filter. We bias subtract and flat field our images using same Python-based routines (PRISM images contain negligible dark current) as with the McD photometry, before performing PSF photometry using {\tt hipercam} and extracting the optimal light curves with \code{phot2lc}. 

{\bf Lowell Discovery Telescope (LDT)} --- We observed \obj\ for 1.82 hours on 2024~August~25 using the Large Monolithic Imager (LMI) at the 4.3-m Lowell Discovery Telescope in Happy Jack, Arizona. To reduce readout times to $\approx$\,4.2\,s we windowed the array to 500$\times$500\,pixels to achieve a 2.4\arcmin$\times$2.4\arcmin field of view with 3$\times$3 pixel binning (0.36\arcsec per pixel). We integrated the field with 15\,s exposures in the SDSS $g$-band. These photometry were reduced in the same manner as the McD and PTO photometry, again using {\tt hipercam}, following bias subtraction and flat fielding. The optimal light curve was extracted using {\tt phot2lc}.

\subsection{Survey Photometry}\label{sec:survey_phot}

{\bf Zwicky Transient Facility (ZTF)} --- We obtained photometry from the Zwicky Transient Facility \citep{Bellm2019,Masci2019} using the ZTF Forced Photometry Service \citep{Masci2023} in identical fashion to \citet{Hermes2025}. Following quality filtering we maintain 1163 epochs in $g$ and 1472 in $r$ spanning 2018~March through 2025~July. We disregard the $i$-band light curve due to its relative paucity of epochs and diminished S/N.

Due to the faintness of \obj\ ($G=19.37$\,mag) and the sufficiently dense sampling by ZTF, we elect against querying other time-domain surveys. We further reject analyzing the NEOWISE light curve of \obj, since every observation from the PSF light curve generated following the method of \citet{Guidry2024} fails quality vetting.

\subsection{Spectroscopy}\label{sec:spectroscopy}

Over four separate nights we obtained 13 spectra of \obj\ using the Double Spectrograph (DBSP, \citealt{Oke82}) on the Palomar 200-in Hale Telescope, and the Low Resolution Imaging Spectrograph (LRIS, \citealt{Oke95,Rockosi10}) on the Keck-I 10-m telescope at Keck Observatory. Eight of these spectra have been previously reported in \citet{Bhattacharjee2025}, while here we report five new LRIS spectra obtained concurrently with CHIMERA time-series photometry. A summary of both the new and previously reported spectra can be found in Appendix~\ref{sec:spec_appendix}.

CHIMERA observations from 2023~June~19 showed three distinct transit features and confirmed the detection of a roughly 5\,hr period from 2023~May observations. We were able to predict the timing of the same three transit features during the following night of 2023~June~20. Concurrently with CHIMERA $g+r$-band observations, we obtained five consecutive 600\,s LRIS exposures on 2023~June~20 that overlapped with the deepest and longest lasting of these transits. As a result, the first three spectra occurred in-transit at average transit depths of around 20\%, while the last two spectra occurred out-of-transit (see Figure~\ref{fig:EW_transit}). On the blue arm we used the 600 line~mm$^{-1}$ grism blazed at 4000\,\AA, and on the red arm we used the 600 line~mm$^{-1}$ grating blazed at 7500\,\AA. With this setup we achieve resolving powers of $R=1100$ on the blue side, and $R=1550$ on the red side for the 400 and 600~line~mm$^{-1}$ gratings, respectively. These setups provide continuous spectral coverage of $3140$$-$$8820$\,\AA\xspace with the split between blue and red arms occurring at $\lambda_{\mathrm{s}}=5644$\,\AA\xspace. We reduced the spectra using the LRIS automated reduction pipeline (LPipe, \citealt{Perley19}). We chose the standard stars Feige\,34 and Feige\,67 for flux calibration.

\section{Recurrent 4.97-Hour Transiting Debris}\label{sec:period}

Below we detail our methods for determining the 4.9704\,hr orbital period of the transiting debris at \obj. We period search only the photometry from our follow-up that includes consecutive nights and still show transits (2023~May -- 2023~November; Figure~\ref{fig:phasedLCs_22-23}; Appendix~\ref{sec:ts_phot_appendix}). We also search the entire ZTF light curve for periods.

\subsection{Lomb-Scargle Periodogram}

\begin{figure*}[t!]
    \epsscale{1.18}
    \plotone{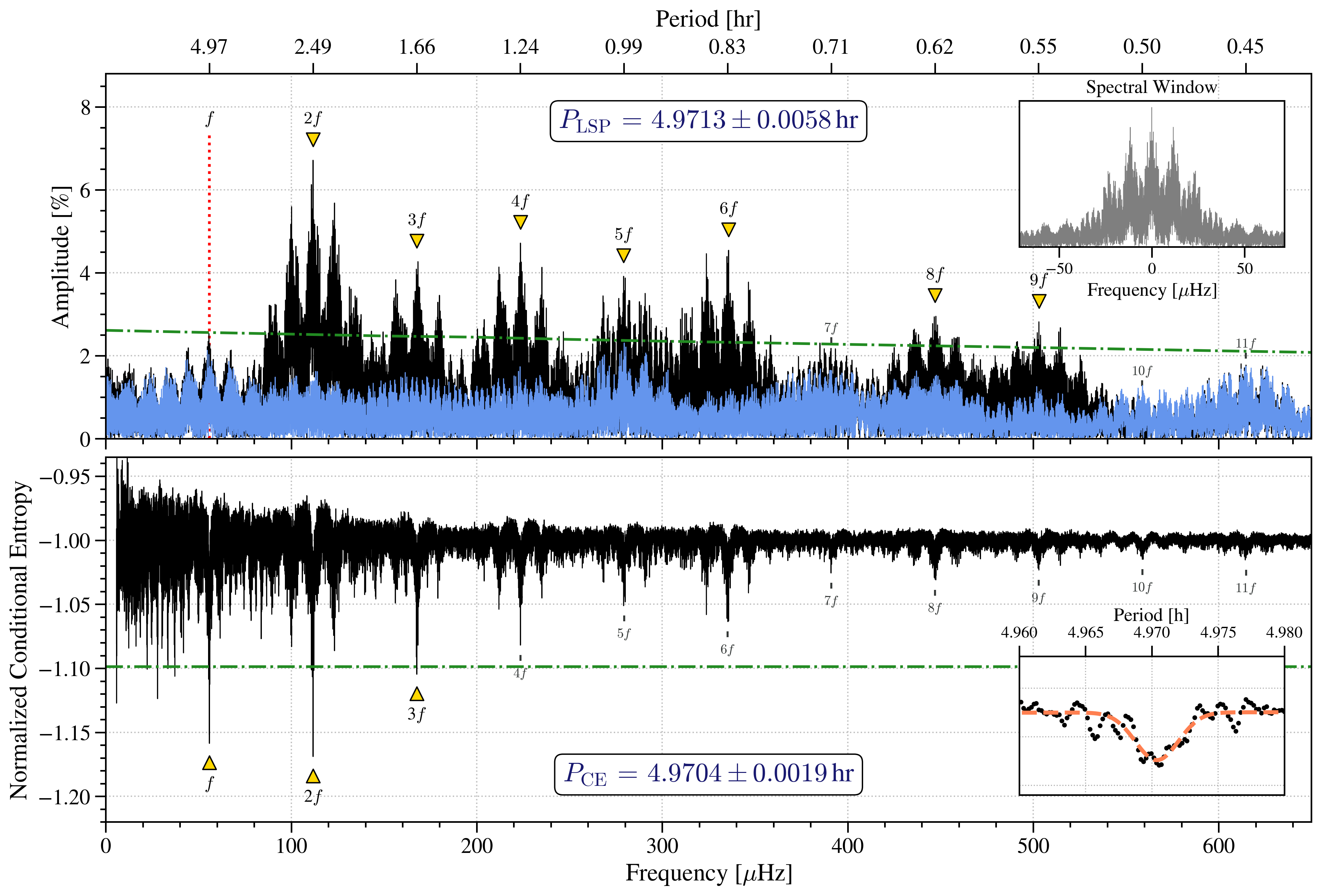}
    \caption{Top: The Lomb-Scargle periodogram of all non-overlapping 2023 CHIMERA-$g$, McD-$BG40$, and PTO-$BG40$ observations. Black shows the raw periodogram while blue shows the residual periodogram after prewhitening the peaks detected in excess of the $5 \langle A \rangle$ threshold (the green dash-dot line, where $\langle A \rangle$ is the mean of the residual periodogram within a sliding window of width 2000\,$\mu$Hz), indicated by yellow triangles. The frequency corresponding to the orbital period of 4.9713\,hr ($f$) is indicated by a vertical red dotted line. High-harmonic power is present given the non-sinusoidal nature of the transits, and we label up to the 11th harmonic ($11f$) even if the peaks fall below the $5\langle A \rangle$ threshold. The period in hours is given along the top axis for each labeled peak, while an inset plot shows the spectral window of the observations. Bottom: Conditional entropy periodogram for the same non-overlapping 2023 CHIMERA-$g$, McD-$BG40$, and PTO-$BG40$ observations, set on the same frequency and period scales as the above Lomb-Scargle periodogram. A $\rho = 10$ threshold (see \citealt{Katz2021}) is plotted as the dash-dot green line. Harmonics of the fundamental are labeled in the same fashion as above. We inset the Gaussian fit that yields the annotated best period of $P=4.9704\pm0.0019$\,hr. \label{fig:LSP_CE_2023}}
\end{figure*}

Given the distinct and unique shapes of the transit features in \obj, we initially were able to gain a rough idea of the debris orbital period by matching up transit features in the 2023 May and June CHIMERA observations by eye, suggesting an orbital period around 5\,hr. To measure the period, we use Astropy routines to calculate unnormalized\footnote{See ``PSD normalization'' at the \href{https://docs.astropy.org/en/stable/timeseries/lombscargle.html}{Astropy Website}} Lomb-Scargle periodograms \citep{Lomb1976,Scargle1982}, which we convert from power ($p$) to fractional amplitude units ($A$) with the relation $A=\sqrt{4p/N}$, where $N$ is the total number of light curve data points. To calculate the periodogram, we only use non-overlapping CHIMERA-$g$, McD-$BG40$, and PTO-$BG40$ observations from 2023~May to 2023~November, prioritizing CHIMERA-$g$ when datasets overlapped due to its higher signal-to-noise and temporal resolution. The table in Appendix~\ref{sec:ts_phot_appendix} indicates the exact datasets used, and the resulting periodogram is shown in black in Figure~\ref{fig:LSP_CE_2023}.

To identify prominent peaks, we use the \texttt{Pyriod} Python package \citep{Pyriod2022} and estimate a 0.1\% false-alarm probability level using five times the mean periodogram amplitude within a sliding window of width 2000\,$\mu$Hz \citep{Breger1993,Kuschnig1997}\footnote{We emphasize a $5\langle A \rangle$ threshold is only correlated with formal statistical significance, and should not be equated with a rigorously determined false-alarm probability threshold. We direct readers to the following for a priming overview on the statistics of periodicity significance: \citet{Baluev2008}; \citet{Hara&Ford2023}, and the references therein.}, shown by the green dash-dotted line in Figure~\ref{fig:LSP_CE_2023}. We accept all peaks above this threshold, and we optimize these periods using \texttt{LMFIT} \citep{LMFIT_2014} by performing a non-linear least squares fit of a sinusoid at each period to the time-series photometry. We then undertake an iterative prewhitening procedure, subtracting the best-fit sinusoids from the time series photometry and calculating a new periodogram and $5 \langle A \rangle$ threshold of the residuals and again look for any new peaks. We repeat this process until no more peaks are found above the $5 \langle A \rangle$ threshold.

This prewhitening procedure results in the detection of seven $>$$5 \langle A \rangle$ peaks, marked with yellow triangles in Figure~\ref{fig:LSP_CE_2023}, all of which are harmonics of the likely orbital period. The base orbital period does not surpass $5 \langle A \rangle$ in the Lomb-Scargle periodogram, however, perhaps due to the highly non-sinusoidal nature of the light curve. The highest amplitude peak at a period of 2.49\,hr corresponds to the second harmonic of the true orbital period. When refining the periods with \texttt{Pyriod}, we constrain all other peaks to be exact harmonic multiples of the highest peak. By doubling the period of the highest peak, we obtain $4.9713 \pm 0.0058$\,hr as the debris orbital period. Our reported uncertainty is simply the $1/T$ frequency resolution, where $T$ is the total time baseline covered by the light curves used to calculate the periodograms. We conservatively adopt this value over the smaller estimated uncertainty of $0.0006$\,hr found using the formalisms developed by \citet{Montgomery1999}. The least squares fit similarly provides a much smaller and likely underestimated statistical uncertainty of 0.000003\,hr.

\subsubsection{In-Transit Harmonics}\label{sec:transit_harmonics}

Inspired by the significant power observed at high-order harmonics of the orbital periods of other transiting debris systems, especially the 65:1 resonance at WD\,1054$-$226 \citep{Farihi2022}, we partition the transits from a sample of 2023 light curves to compare the Lomb-Scargle periodogram of the in-transit variations to that of the aggregate, full light curve in Figure~\ref{fig:LSP_CE_2023}. We select the CHIMERA $g$-band light curves from 2023~May~21 \& 25, June~19 \& 20, and August~12, and the 2023~September~20 ProEM light curve. We identify transits by searching for groupings of observations whose fluxes are less than the difference between a given light curve's median flux and its median absolute deviation: $f_{\rm transits} < f_{\rm median} - f_{\rm MAD}$. To be considered real transits, these flux groupings must have durations of at least 10 consecutive exposures. Nearby groups will be consolidated if they are separated by no more than 600\,s. The potential transits that pass these restrictions must not be partial: they must include a fully resolved ingress and egress. We visually vet all those that pass these criteria, confirming they are real transits, which generally include the transit ingresses and egresses. We reserve the timestamps and fluxes of all detected transits, record the barycentric-correct Julian Date of each first observation, and aggregate them into a single, combined, in-transit-only light curve. We re-normalize the relative in-transit fluxes by adding the difference between 1.0 and the median flux of a given transit so that the in-transit variations are all the same relative flux scale.

We compute the Lomb-Scargle periodogram, oversampled by a factor of 30, of the in-transit photometry to compare against the non-partitioned photometry. The large data gaps between transits within this light curve complicate standard techniques for bootstrapping significance thresholds. This in-transit periodogram appears dominated by peaks at nearly every harmonic of the fundamental 4.9704\,hr (55.8864\,$\mu$Hz) orbital period starting at its fifth harmonic. We identify the following peaks in the periodogram that exceed 3\% in amplitude (4.25\,$\langle A \rangle$): 449.0477\,$\mu$Hz (3.05\%), 615.6358\,$\mu$Hz (3.27\%), 839.7629\,$\mu$Hz (3.15\%), 949.6308\,$\mu$Hz (3.21\%), 1061.4114\,$\mu$Hz (3.21\%), 2121.4981\,$\mu$Hz (3.32\%), 2232.3903\,$\mu$Hz (3.31\%). These could correspond to the 8th, 11th, 15th, 17th, 19th, 38th, and 40th harmonics, respectively, of the 4.9704\,hr (55.8864\,$\mu$Hz) orbital period. There are 35 peaks that exceed an amplitude of 2\% (2.83\,$\langle A \rangle$), among which the highest in frequency (3852.9670\,$\mu$Hz, 2.17\%) could correspond to the 69th harmonic. That these high-order harmonics are observed in-transit suggests that there exists ordered structure to the transiting material, evoking the numerous high harmonics found in the TESS photometry of WD\,1054$-$226 \citep{Farihi2022}, particularly the 65:1 orbital resonance.

\subsection{Conditional Entropy}

We further period search the same set of 2023 light curves light curves using conditional entropy \citep[][]{Graham2013_CE} as implemented by {\tt cuvarbase}\footnote{\url{https://johnh2o2.github.io/cuvarbase/}} \citep{cuvarbase2022}. Unlike the Lomb-Scargle method, conditional entropy is agnostic to the shape of the periodic variability and incorporates the uncertainties on the the time series as it searches the trial frequency that minimizes disorder across the phase-magnitude grid relative to the un-phased light curve. Conditional entropy further safeguards against aliasing from observing cadences. For transiting debris that is highly non-sinusoidal, such as the features in Figure~\ref{fig:phasedLCs_22-23} and those seen at ZTF\,J0328$-$1219 and WD\,1054$-$226 \citep{Vanderbosch2021,Farihi2022}, and irregularly sampled, we suggest conditional entropy is optimal for period searching.

We reuse the grid of trial frequencies from our Lomb-Scargle periodogram, now oversampled by a factor of 30 (as done by \citealt{El-Badry2022}), and compute conditional entropy using 20 phase bins and 10 magnitude bins (we do not convert our relative fluxes to magnitudes). Before searching for the minimum we subtract a smoothed periodogram obtained by fitting a sliding median filter using {\tt scipy} to the periodogram with a width of 17.35$\,\mu$Hz. This removes systematics at low frequencies with the added benefit of effectively normalizing the periodogram \citep{Graham2013_CE}.

We find the conditional entropy to be minimized at $P = 2.4582$\,hr, at an equivalent amplitude of $\rho = 17.2$\footnote{We use $\rho$ (see Equation~5 of \citealt{Katz2021}) here to quantify how many standard deviations below the mean a given local minimum is in the periodogram as a gauge for significance. Like \citet{Katz2021}, we adopt $\rho = 10$ as a threshold for peaks likely attributable to astrophysical variability.}, with the second-lowest minimum at about half this period, $P = 4.9705$\,hr at $\rho = 16.1$. To refine this measurement, we fit a Gaussian function using {\tt LMFIT} centered on the core of the $P = 4.9705$\,hr minimum in the CE periodogram (see the inset in the lower panel of Figure~\ref{fig:LSP_CE_2023}). Here we estimate uncertainty on the mean of the function from the half width at half maximum, finding a best-fit period of $P_{\rm orb} = 4.9704 \pm 0.0019$\,hr. This is the measurement we adopt as the orbital period of the transiting debris at \obj, again since conditional entropy is agnostic to the shape of the model of this non-sinusoidal variability.

When applying the same procedure to the combined $g+r$ ZTF light curve, now normalized to a median filter with a width of 2.26\,$\mu$Hz, we measure a periodicity at 4.9703\,hr at an amplitude of $\rho =  4.75$. All other possible periodicities, none of which clear $\rho = 10$, appear to be driven by the near-diurnal survey cadence.

\section{In-Transit Spectroscopy} \label{sec:pollution}

We acquired spectra of \obj\ on four separate nights, initially using the DBSP spectrograph on the Palomar 200-in telescope on 2022~May~6. These observations indicated the presence of metallic absorption features with clear Ca-H and K lines. We sought higher S/N observation on 2022 July 4 using the LRIS spectrograph on the Keck-I 10-m telescope. These observations confirmed the presence of metallic absorption features at Ca-H and K, but without significant detections of any other metallic species.

Spectra from these first three nights were reported in \citet{Bhattacharjee2025}, and the LRIS spectrum was used to fit atmospheric models to derive the atmospheric parameters: \teff$\,{=}\,20{,}790\pm250\,$K and $\log(g\,[\mathrm{cm\;s^{-2}}])\,{=}\,8.27\pm0.05$, resulting in a white dwarf mass of $M=0.79\pm0.03\,M_{\odot}$. These parameters are in good agreement with the Gaia EDR3 White Dwarf catalog \citep{GentileFusillo2021}, which provides \teff$\,{=}\,20{,}300\pm5400\,$K and $\log(g\,[\mathrm{cm\;s^{-2}}])\,{=}\,8.20\pm0.43$ using H-dominated atmospheric models. 

\citet{Bhattacharjee2025} note, however, that the Ca-H and K lines proved difficult to fit, with models producing much narrower and weaker Ca-H and K lines at this temperature than are observed. They speculate that the Ca-H and K features could be the result of circumstellar absorption, which has been observed in other transiting debris systems \citep[e.g. WD\,1145+017,][]{Xu2016,Fortin-Archambault2020,LeBourdais2024}. They also rule interstellar medium (ISM) absorption as unlikely given the observed strength of the Ca-K line.

\subsection{Equivalent Width Measurements} \label{sec:spec_ew}

\begin{figure}[t!]
    \epsscale{1.2}
    \plotone{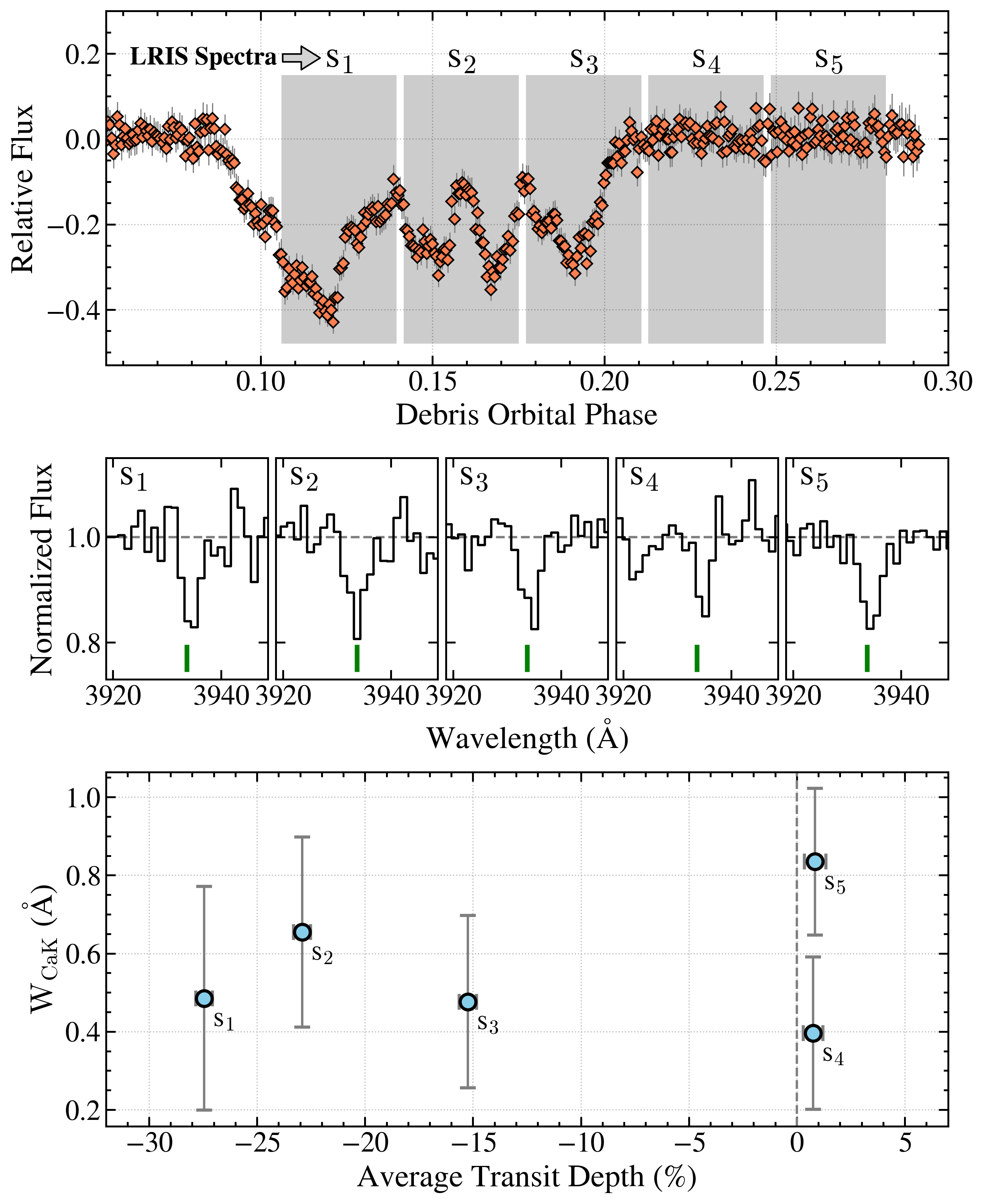}
    \caption{Concurrent time-series photometry and spectroscopy from CHIMERA and LRIS, respectively, during one of the transits observed on 2023~June~20. ({\em Top}) A portion of the 2023~June~20 $r$-band CHIMERA light curve, zoomed in on the transit feature throughout which we acquired five 600-s LRIS exposures, whose time-spans are denoted by the gray shaded regions. ({\em Middle}) Each continuum-normalized LRIS spectrum, centered on the Ca-K absorption feature at 3934\AA. Green lines denote the rest wavelength of Ca-K. ({\em Bottom}) The measured Ca-K equivalent widths versus the average photometric transit depth during each spectrum. Within our measurement uncertainties of around 0.15$-$0.2\AA\ (about 15$-$20\%), the equivalent widths are consistent with being non-variable throughout the transit event. Variations could occur if some portion of the observed absorption were of circumstellar origin, and the transiting debris preferentially obscured a larger portion of the white dwarf than a gas disk interior to the transiting cloud of debris, or vice versa. WD\,1145+017 shows in-transit bluing due to weaker circumstellar gas absorption from such a geometric effect \citep{Hallakoun2017, Karjalainen2019, Xu2019a}.}  \label{fig:EW_transit}
\end{figure}

Following the initial detection of irregular transit features in \obj, We obtained five additional spectra of \obj\ on 2023~June~20 using Keck LRIS, this time concurrently with CHIMERA time-series photometry (Figure~\ref{fig:EW_transit}). These spectra were taken to coincide with a predicted transit feature to look for variations in Ca-K absorption line strength as a function of transit depth. Such correlations, like the observed in-transit weakening of metallic circumstellar absorption lines in WD\,1145+017 \citep{Hallakoun2017,Karjalainen2019,Xu2019a}, could help constrain possible orbital geometries of the transiting debris. The first three spectra occur in-transit, at average depths ranging from about $-27.4\%$ to $-15.5\%$, 
while the last two spectra occurred out-of-transit, at average fluxes of about $0.75\%$ (see Appendix~\ref{sec:spec_appendix}). 

We measure the equivalent width of the Ca-K line by first continuum normalizing each spectrum in the range 3910--3960\,\AA, and then integrating the normalized spectra over a 30\AA\xspace wide region centered on the Ca-K rest wavelength in air of 3933.66\AA. The bottom panel of Figure~\ref{fig:EW_transit} shows the equivalent width measurements versus the average transit depth during each spectroscopic exposure, and the equivalent width values are reported in Appendix~\ref{sec:spec_appendix}.

We do not find any significant correlation between transit depth and Ca-K equivalent width during the LRIS and CHIMERA concurrent observations. The combined in-transit spectrum has \eqwk$\,=0.59\pm0.13$\,\AA, while the combined out-of-transit spectrum has \eqwk$\,=0.63\pm0.14$\,\AA, well within 1$\sigma$ agreement of one another. All five spectra from these concurrent observations have a mean equivalent width of $0.57$\,\AA, with a standard deviation of $0.16$\,\AA. These in-transit, out-of-transit, and averaged values are also within 2$\sigma$ agreement with spectra taken during the first three nights, with Ca-K equivalent widths for the combined spectra of $0.79\pm0.29$\,\AA, $0.91\pm0.16$\,\AA, and $0.99\pm0.29$\,\AA\ for the DBSP 2022~May~6, LRIS 2022~July~4, and DBSP 2022~November~21 observations, respectively. Spectra from these three nights, however, lack constraints on the transit depth at time of exposure. 

We find a mean of all 10 \eqwk\ measurements listed in the table in Appendix~\ref{sec:spec_appendix} of 0.72\,\AA\ with a standard deviation of 0.29\,\AA. Thus, we find the Ca-K equivalent widths to be constant in and out of transit to a limit of at least 0.29\,\AA (40\% fractional uncertainty).

We additionally test for a correlation of the equivalent widths of two more prominent metal features, around 5169\,\AA\ where there could be a blending of several Fe-II and Mg-II transitions and the Na-I D1 (5896\,\AA) and D2 (5890\,\AA) doublet, with the observed concurrent transit depth. Around the 5169\,\AA\ feature, we observe a 2-$\sigma$ tension in the equivalent widths in  and out of transit: the aggregate in-transit spectrum shows \eqwfe$\,= 0.35 \pm 0.22$\,\AA; out of transit, \eqwfe$\,= 0.69 \pm 0.06$\,\AA. At the Na-D doublet, the combined equivalent widths in and out of transit are \eqwna$\,= 0.98 \pm 0.13$\,\AA\ and \eqwna$\,= 0.99 \pm 0.10$\,\AA, respectively.

\subsection{Radial Velocity Measurements} \label{sec:spec_rv}

We also measure the radial velocities (RVs) of four groups of absorption features. The first group includes five H-Balmer lines from H$\beta$ (4861\,\AA) through H8 (3890\,\AA), the second group includes just the Ca-II~K (3934\,\AA) line, the third group includes three Fe-II lines at 4924, 5018, and 5169\,\AA, and the fourth group includes the Na-I~D doublet. We split up the Ca-II , Fe-II, and Na-D lines in case disproportionate levels of photospheric, circumstellar, or ISM absorption, along with potentially flawed sky subtraction, affect each group differently. We fit all lines simultaneously with a combined Voigt profile and linear model, with each line within a group constrained to have the same RV shift. We fit only the LRIS spectra, as the DBSP spectra are significantly lower S/N and often suffer from poor flexure corrections \citep[see Section~3.3 in][]{Nagarajan2023} along with instrumental artifacts at wavelengths ${<}4000$\,\AA\xspace that could skew the velocity measurements. Individual LRIS exposures within a single night do not show any RV variations within line groups ${>}\,2\sigma$ significant, so we report in Table~\ref{tab:RVs} only the RVs from the combined LRIS spectra for the two nights 2022~July~4 and 2023~June~20. We do not detect RV shifts exceeding 2.4$\sigma$ significant within individual line groups between the two epochs.

Given the RV uncertainties and the relatively large RV variations between nights for the H-Balmer lines, we do not find any significant RV differences between H-Balmer and other line groups that would conclusively indicate the presence of ISM or circumstellar absorption in addition to gravitationally redshifted photospheric absorption. Perhaps noteworthy, however, is the relative stability of the Fe-II and Na-D line velocities between nights compared to the Ca-K line velocity. This may suggest the Fe-II lines are mostly the result of photospheric absorption while Ca-K is a blended line more heavily impacted by some time-variable circumstellar absorption, like that observed in WD\,1145+017 \citep{Redfield2017,Fortin-Archambault2020,LeBourdais2024}. The Na-D line velocities are also quite stable between nights, and given their proximity to zero velocity in our LRIS spectra, which have heliocentric velocity corrections applied, these lines may simply be the result of poor sky subtraction.

\begin{deluxetable}{cccc}[!h]
\tablenum{1}
\tablecaption{Keck+LRIS RV Measurements\label{tab:RVs}}
\tabletypesize{\footnotesize}
\tablehead{
    \colhead{Line Group} & \colhead{RV (2022 July 4)} & \colhead{RV (2023 June 20)}\\
    \colhead{} & \colhead{km\,s$^{-1}$} & \colhead{km\,s$^{-1}$}
    }
\startdata
H-Balmer &  $78.7\pm15.1$ & $34.6\pm10.8$\\
Ca-K  & $125.4\pm11.3$ & $90.6\pm10.5$\\
Fe-II    &  $71.4\pm12.0$ & $69.1\pm14.0$\\
Na-D    &  $-6.9\pm14.8$ &  $4.0\pm16.0$\\
\enddata
\tablecomments{The line groupings are the five Balmer lines from H$\beta$ -- H8, the Ca-K line, the three Fe-II lines at 4924, 5018, and 5169\,\AA, the Na-I~D doublet.}
\vspace{-1.5cm}
\end{deluxetable}

\vspace{-0.5cm}
\subsection{Evidence for Circumstellar Gas} \label{sec:CS_gas}

\begin{figure}[t!]
    \epsscale{1.18}
    \plotone{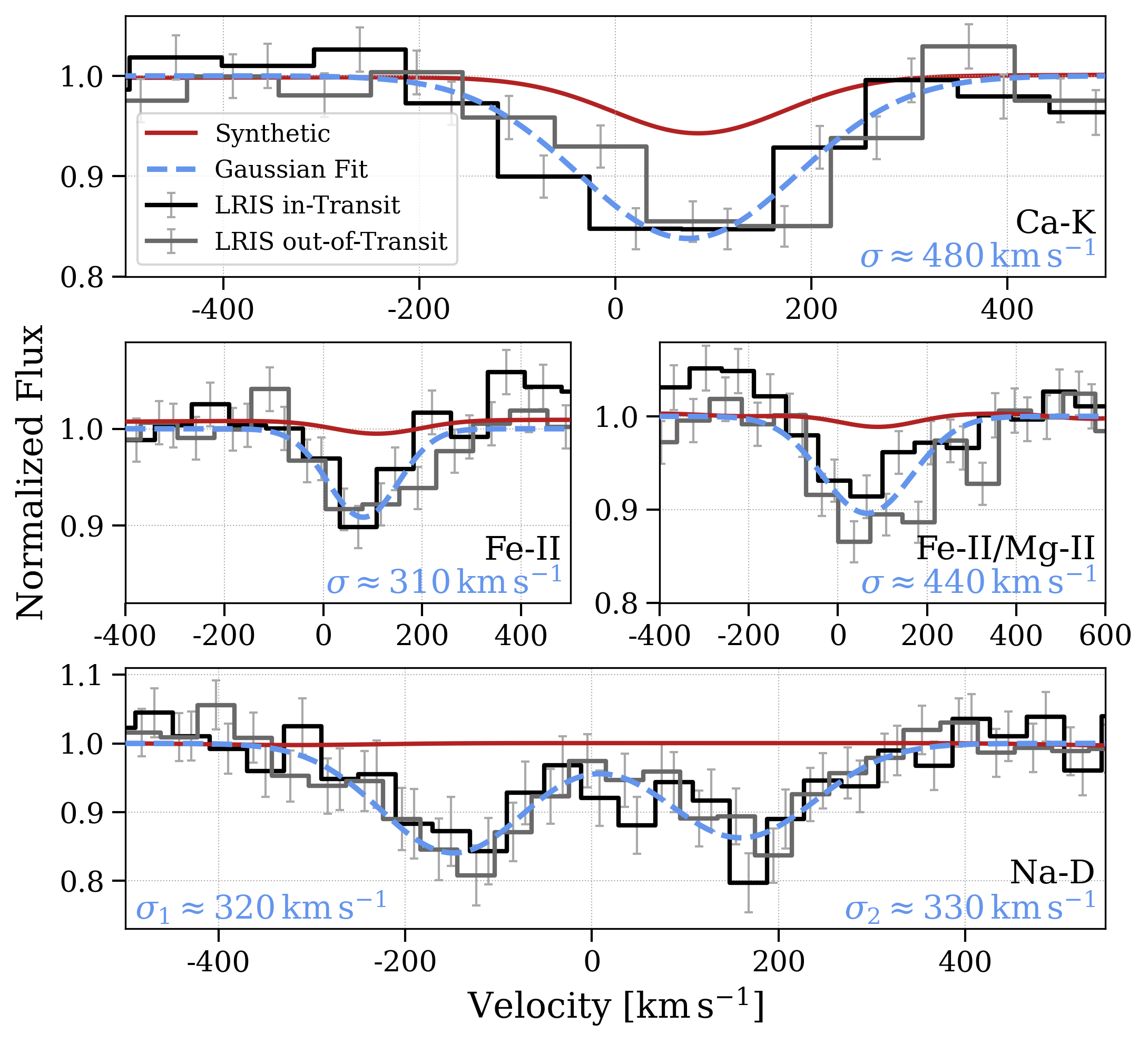}
    \caption{Magnifications of the Ca-K (top), Fe-II 5018\,\AA\ (center left), 5169\,\AA, which could be a blend of Fe-II and Mg-II lines (center right), and Na-D doublet (bottom) transitions as observed by our 2023~June~20 Keck+LRIS spectra and modeled by our synthetic DAZ spectrum, all converted to velocity space. We aggregate all five spectra from Figure~\ref{fig:EW_transit} and fit Gaussian profiles to the cores and report the associated velocity dispersions. Our synthetic spectrum is generated for a $T_{\rm eff} = 20{,}794$\,K, $\log(g\,[{\rm cgs}]) = 8.276$ DA white dwarf accreting bulk Earth material scaled to $[{\rm Ca/H}] = -5.0$, convolved to the LRIS instrumental resolution ($R\approx1500$). Our synthetic spectrum fails to reproduce the observed equivalent widths of these transitions by photospheric absorption alone, suggesting circumstellar gas could be casting these features, especially the Fe-II and Na-D absorption.}\label{fig:spec_zooms}
\end{figure}

While we do not measure spectroscopic variations from in- to out-of-transit (Figure~\ref{fig:EW_transit}), we still find evidence for circumstellar gas in our spectroscopy. Building on the argument by \citet{Bhattacharjee2025} that circumstellar gas is needed to explain the broadness of the observed Ca-K absorption, we magnify in Figure~\ref{fig:spec_zooms} our 2023~June~20 LRIS spectra around the Ca-K line, Fe-II 5018\,\AA\ line, the multiple Fe-II and Mg-II lines around 5169 \AA\ (we are unable to identify which at this resolution), and the Na-D doublet transitions. There, we compare the aggregate in-transit and out-of-transit spectra to a synthetic DAZ white dwarf spectrum \citep{Koester2010} computed using $T_{\rm eff} = 20{,}794$\,K, $\log(g\,[{\rm cgs}]) = 8.276$ and accreting material with a bulk Earth composition scaled to $[{\rm Ca/H}] = -5.0$, one dex further enriched than explored by \citet{Bhattacharjee2025} and generally high for polluted white dwarfs \citep[see, e.g.,][]{Gaensicke2012,Xu2014,Koester2014,Xu2019b,Bonsor2020,Rogers2024}. 

Even with a heavily enriched photosphere, we are unable to reproduce the observed absorption at all four regions in Figure~\ref{fig:spec_zooms}. The presence of the Fe-II and Na-D lines are of particular note. The line ratios of the observed Fe-II lines are consistent with a temperature far cooler than the photospheric temperature ($T_{\rm eff} \approx 10{,}000$\,K, for example, better matches the observed line strengths); this gaseous iron is likely external to the white dwarf. Likewise, sodium would be ionized at $T_{\rm eff} \approx 20{,}000$\,K, and so the observed Na-D absorption cannot be photospheric in origin. Though imprecise, the velocities of the Ca-K and Fe-II lines from both of our LRIS epochs are removed from the heliocentric velocities of those nights by over $4\sigma$, suggesting they are likely not an artifact of sub-standard sky subtraction. 

Interstellar gas is not predicted to produce such broad features like those in our spectroscopy. \eqwk$ = 0.72$\,\AA\ is towards the upper-end values from pure ISM absorption that have been observed at distances ${<}\,500\,$pc \citep{Welsh2010}; \obj\ lies at a distance of $447^{+44}_{-40}\,$pc \citep{BailerJones2021} from Earth. \obj is located at $l=10.71$\arcdeg, elevating it to $Z \approx 100$\,pc above the galactic plane, where it is estimated that the ISM contributes \eqwk$ \approx 0.02$\,\AA\ \citep{Beers1990}. More recent surveys indicate \eqwk$ \approx 0.3$\,\AA\ at $l\approx10$\arcdeg. The line-of-sight extinction towards \obj\ is not particularly high, with 3-dimensional extinction values of $A_0=0.14$\,mag from Gaia-2MASS \citep{Vergely2022}, $A_V=0.14$\,mag from Bayestar19 \citep{Green2015,Green2018}, and $A_V=0.15$\,mag from the Gaia eDR3 white dwarf catalog \citep{GentileFusillo2021} derived using the extinction maps built by \citet{Vergely2022}. \citet{Bhattacharjee2025} derived $A_V=0.192$\,mag, for comparison. At a color index of $E(B-V)=0.06$ \citep{Bhattacharjee2025}, interstellar calcium is expected to cast an equivalent width of \eqwk$\,\approx 0.3$\,\AA\ \citep{Murga2015}, also over a factor of 2 less than our measurement. Sodium tells a similar story. The same color index correlates with ${\rm W_{D_1+D_2}} = 0.54 \pm 0.07$\,\AA\ \citep{Poznanski2012} (\citealt{Murga2015} find ${\rm W_{D_1+D_2}} \approx 0.5$\,\AA), yet we measure${\rm W_{D_1+D_2}} = 1.6 \pm 0.3$\,\AA\ in the combined 2022~July~04 LRIS spectrum and on 2023~June~20 ${\rm W_{D_1+D_2}} = 1.0 \pm 0.1$\,\AA. Circumstellar gas appears to best explain the metal enrichment of our spectroscopy, but it is plausible there are also contributions from the ISM.

Higher-resolution spectroscopy is necessary to constrain any line strength variability and potentially  resolve individual circumstellar, photospheric, and ISM absorption components in \obj \citep[e.g.][]{Debes2012}. The faintness of \obj\ ($G=19.4\,$mag) complicates high-resolution optical observations. Far-ultraviolet spectroscopy could be more viable, as the SED of \obj peaks in the far-UV and the heightened sensitivity to metal transitions in polluted white dwarfs in the UV \citep[e.g.,][]{Gaensicke2012}. This is a viable option for \obj given its high effective temperature making it considerably brighter in the far-UV than at optical wavelengths.

The potential gas disk at \obj would likely be composed of concentric eccentric rings \citep{Cauley2018,Fortin-Archambault2020,LeBourdais2024}, all arrayed interior to the 4.9704-hr orbit from the constraint of the equilibrium and sublimation temperatures (Section~\ref{sec:orbit}). We see in Figure~\ref{fig:spec_zooms} broad velocity dispersions reminiscent of the circumstellar gas at WD\,1145+017, which shows asymmetric velocity profiles with total dispersions of about 300\,km\,s$^{-1}$ \citep{Xu2016}. While the features towards \obj appear symmetric at the LRIS resolution limit in Figure~\ref{fig:spec_zooms}, all four lines are consistent with dispersions exceeding 300\,km\,s$^{-1}$. We estimate these velocity dispersions from the width between the points where the amplitude of the Gaussian fits reach their upper 10th percentile. 

If confirmed, \obj would join WD\,1145+017, ZTF\,J0328$-$1219, and WD\,J1013$-$0427 \citep{Bhattacharjee2025} as the only four out of the to-date 14 published transiting white dwarfs to show circumstellar gaseous debris either in absorption or emission (ZTF\,J0139+5245 shows enhanced absorption at Ca-K during its deep transits, possibly due to an augmentation in the column density of gaseous debris along the line of sight; see \citealt{Vanderbosch2020}).

\section{Observed Properties of the Transits at \obj} \label{sec:discussion}

We dedicate this section to surveying the observed and measured properties of the periodic transits at \obj and place them into the broader context of this emerging class and the predicted theory for these systems.

\subsection{Variable Transit Activity}

The constantly changing transit morphologies in Figure~\ref{fig:phasedLCs_22-23} indicate rapid circumstellar dynamical timescales at \obj. If we assume our best-determined 4.9704\,hr period to be the true orbital period of the transiting debris, then dynamical processes, particularly collisions within the debris, must occur on a timescale $t_{\rm collision} < P_{\rm orb}$. Ongoing collisional activity and Keplerian shear could explain why no single transit feature appears to repeat identically from one cycle to the next. This is in contrast to white dwarfs like WD\,1054$-$226 with 25.02-hr recurring transit features that remain stable over a timescale of weeks \citep{Farihi2022}, and raises challenges to mapping the phase migration of specific features as done at WD\,1145+107 \citep{Gaensicke2016,Rappaport2016,Gary2017,Aungwerojwit2024}.

\begin{figure}[t!]
    \epsscale{1.18}
    \plotone{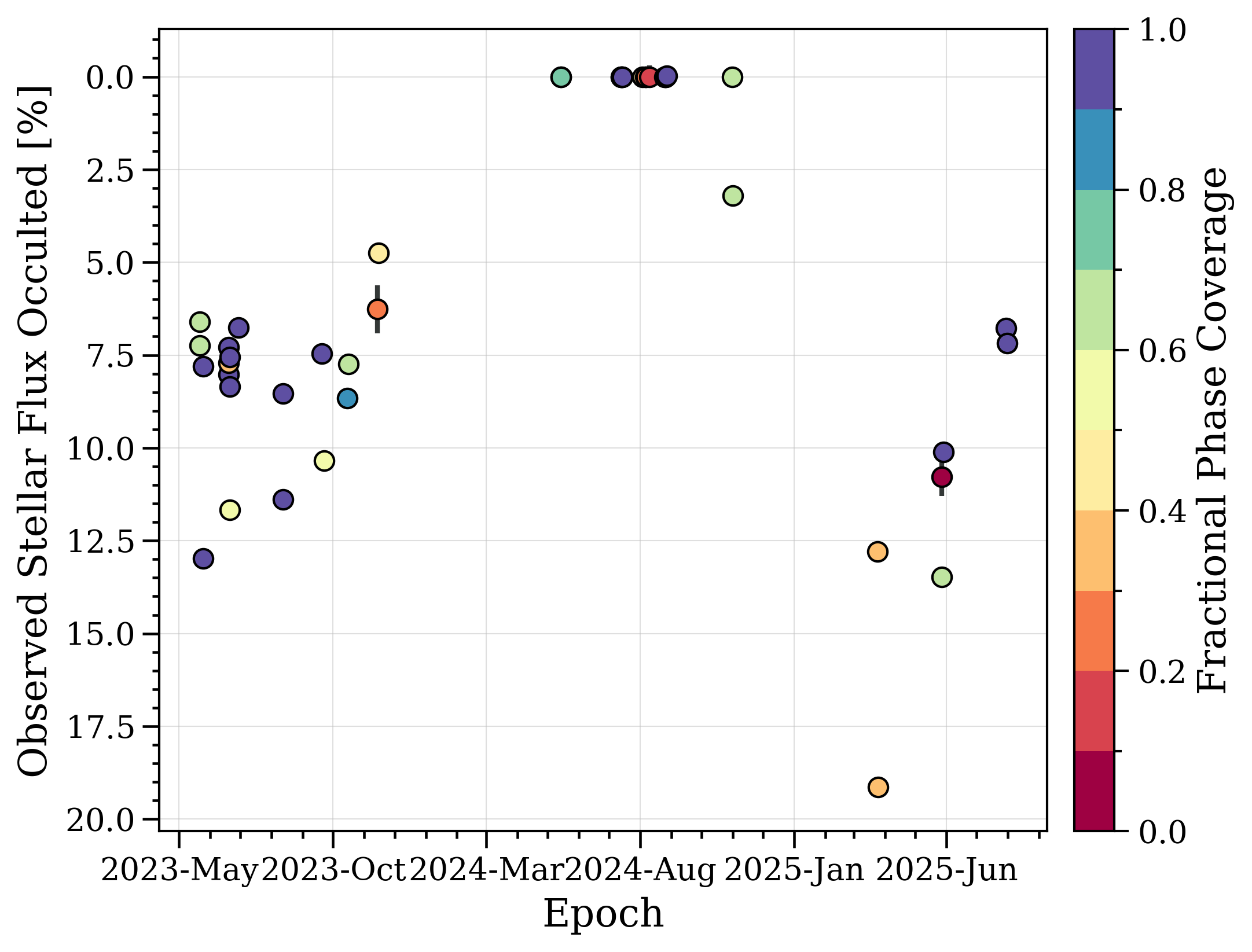}
    \caption{The total transit activity level measured as the percent of white dwarf flux blocked along the line of sight during transit events relative to the duration of the light curve. We exclude light curves from 2022 because of their relatively short baselines and minimal transit activity. All light curves fluxes except those from 2024~May--October are shifted down by the mean of the flux values greater than the 80th percentile to compensate for deep transits skewing the out-of-transit continuum from zero when normalizing the light curve during our reduction process. Marker colors map to the fractional phase coverage of the light curve ($P_{\rm orb} = 4.9704\,$hr). At the highest activity levels observed, just over 19\% of white dwarf flux is blocked along our line of sight during transit events, while at the lowest activity levels, few if any transits are observed (see also Figures~\ref{fig:phasedLCs_22-23} and \ref{fig:phasedLCs_24}).}  \label{fig:activity}
\end{figure}

\begin{figure*}[t!]
    \epsscale{1.17}
    \plotone{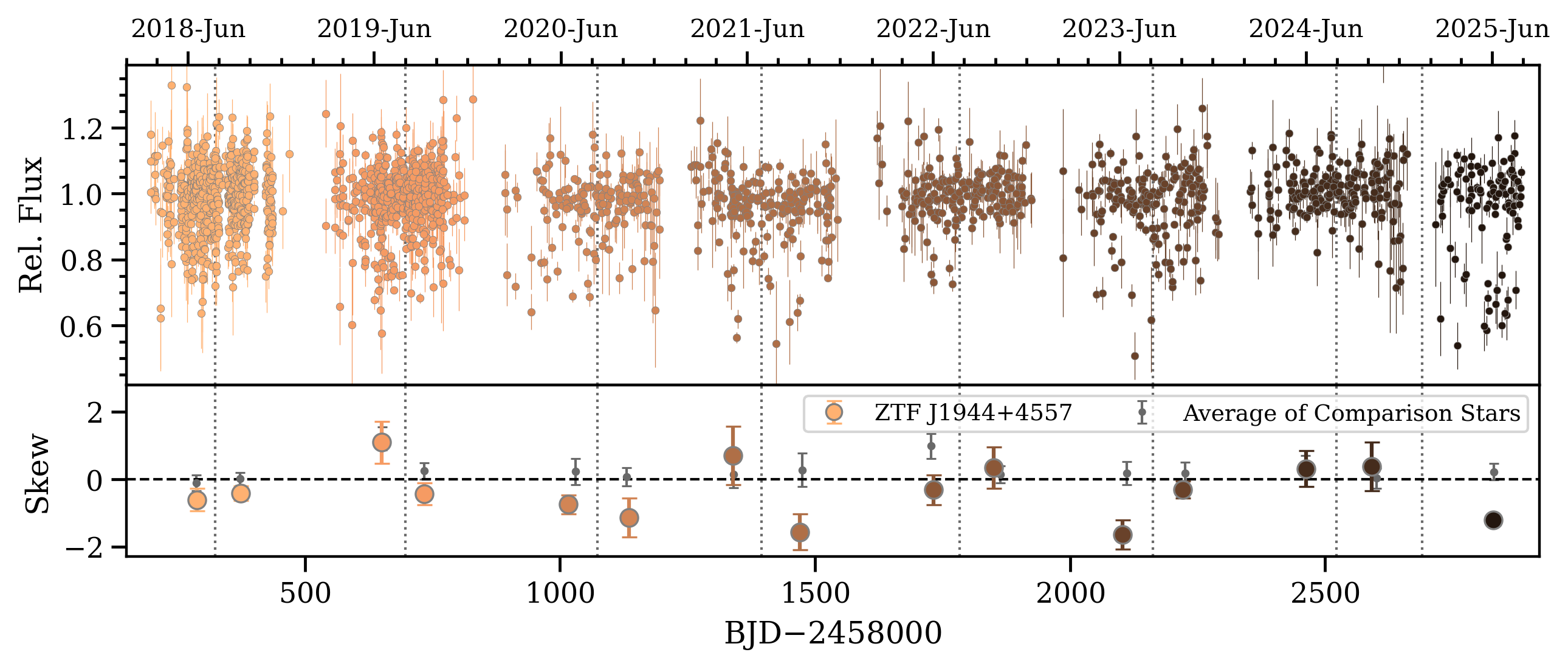}
    \caption{Top: The ZTF forced-photometry light curve of \obj\ partitioned into eight seasons, distinguished by shading. We draw dotted silver-colored lines at the median timestamp of each season, across which we evaluate the skewness (bottom) for each half season; the eighth season is not halved for its shortness. The skew uncertainties are taken as the standard deviation of the skew values found from a Monte Carlo simulation with $10{,}000$ draws for each half season. The smaller gray points are the averages of the same calculations for three nearby non-variable white dwarfs of the same magnitude to provide a benchmark for expected seasonal flux variations. We find a general agreement with our follow-up photometry (Figure~\ref{fig:activity}) that the transits were at heightened activity during the summer of 2023, ceased during the summer and fall of 2024, and recurred in 2025. This could suggest that the transit activity at \obj\ is variable over years-long timescales, fluctuating between heightened stages of transit activity (e.g., Fall 2021) and quiescent periods (e.g., Fall 2022). All half-seasons are consistent with transiting activity or quiescence, with the in-transit seasons showing much larger dispersion than observed for the non-variable reference stars.
    }  \label{fig:ztf_season_skews}
\end{figure*}

\begin{figure*}[t!]
    \epsscale{1.16}
    \plotone{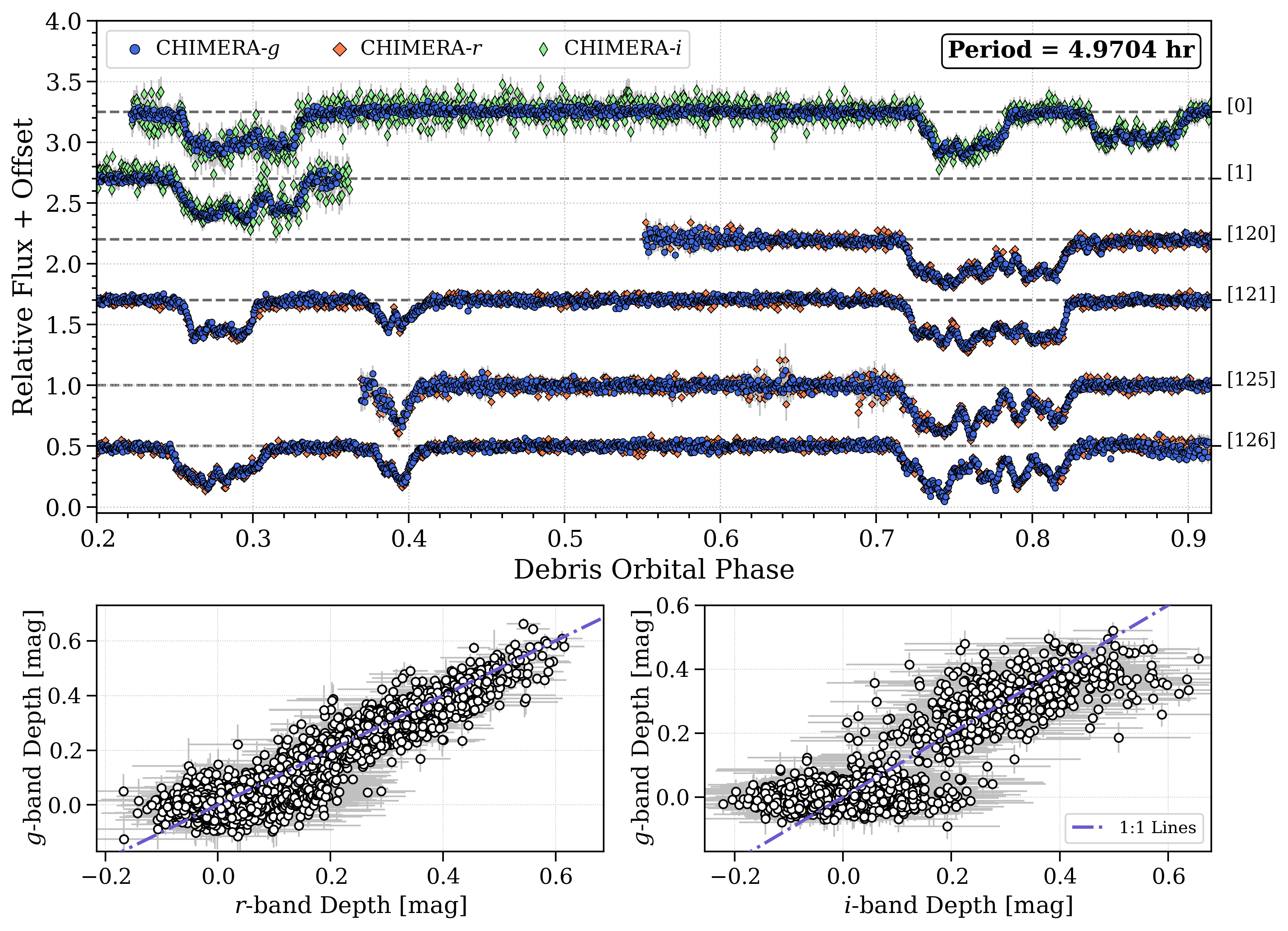}
    \caption{Top: 2023 May 25 $g$+$i$ and 2023 June 19 and 20 $g$+$r$ light curves taken with CHIMERA shown in the same style as Figure~\ref{fig:phasedLCs_22-23}, now magnified and un-binned to better appreciate the rapid variation in transit morphologies and lack of observed color terms. Bottom: Comparison between the transit depths in the $g$-band vs. either the $r$- (left) or $i$-band depths (right), all in differential magnitudes, for our CHIMERA light curves from 2023~May, June, and August. The overlap with the dashed-dotted purple lines that trace the 1:1 relationships suggests the transits at \obj lack color variations (i.e. gray). The transiting effluents appear devoid of small ($\lesssim 0.2\,\mu$m) dust particles. This is rigorously supported by our finding that the extinction coefficients are consistent with zero (Section~\ref{sec:color}).
    }  \label{fig:color_dependence}
\end{figure*}

We attempt to quantify the activity levels of the transit at every epoch we observed \obj\ in Figure~\ref{fig:activity}, where we plot the integrated flux relative to the baseline of our observations at a given epoch \citep[c.f.,][]{Gary2017}. The highest transit activity we observe equates to a just-over 19\% occultation of stellar flux across a full 4.9704-hr orbit. Notwithstanding our occasional incomplete sampling of the orbital cycle, the vanishing of transits from 2024~May--2024~October is obvious. This behavior appears correlated with our estimated activity of the ZTF light curve in Figure~\ref{fig:ztf_season_skews}. There, we partition the over six-years-long light curve into eight seasons. We calculate the Fisher-Pearson skew\footnote{\url{https://docs.scipy.org/doc/scipy/reference/generated/scipy.stats.skew.html}} for first and second halves of each season. We estimate uncertainties on these values by taking the standard deviation of the results of a Monte Carlo simulation where we perturb a given half season's flux over $10{,}000$ draws. 

The season centered around 2023~June shows the most negative skew, matching the elevated degree of transit activity measured by our follow-up during this period. This deviates from the typical scatter for non-variable white dwarfs at this magnitude. The smaller gray points in the bottom panel are the average of three nearby (within 1.2\,deg) high-probability ($P_{\rm WD} > 0.75$; \citealt{GentileFusillo2021}) white dwarfs\footnote{Gaia DR3 source IDs: 2079934773192051712, 2080241850467748864, 2080503220704806528} with $19.0 < G < 19.5$\,mag that are not photometrically variable ({\sc varindex}$_{\rm eDR3}<0$; \citealt{Guidry2021}). The 2024 seasons are consistent with zero skew and returns to a negative skew in early 2025, again agreeing with the cessation and resumption of the transits we observe in our follow-up. The variation in the skew over these six years suggests the transit activity at \obj fluctuates over years-long timescales. The three strongly negative half seasons at approximate BJD$-$2${,}$458${,}$000 of 1500, 2100, and 2700 are each separated by $\approx 635$\,days, or 1.7\,years. Though not yet significant, this evidence for possible secular recurrence is noteworthy for future monitoring.

Our observations of \obj are the second case of a cessation of transit activity at a white dwarf remnant planetary system, and perhaps the first clear resumption of periodic transits. Dense, long-term monitoring of WD\,1145+017 tracks phases of minimal and inflated transit activity \citep{Gaensicke2016,Gary2017}, with transits vanishing in 2022 \citep{Aungwerojwit2024}; their return has yet to be documented. Elsewhere, the 11.2-hr ``B" Period at ZTF\,J0328$-$1219  disappeared in the second half of TESS Sector 31 photometry, whereas the predominant 9.937-hr ``A" period persists across both halves \citep{Vanderbosch2021}.

While we elect to reserve speculation and physical modeling of the vanishing and return of transits to future studies, it merits underlining the physical implications of this result. Poynting-Robertson drag is expected to exhaust typical debris disks at white dwarfs within timescales of $10^4 \lesssim t_{\rm disk} \lesssim 10^6$\,yr \citep{Girven2012,Cunningham2021}. Following the work of \citet{Veras&Hend2020} for tidally disrupted debris disks, we can estimate that a cessation timescale $\sim\,1-2$\,yr generally requires minimum particle sizes $s \gtrsim 10^{-1}$\,m for disks with masses $M \gtrsim 10^{17}$\,g. Transits appear to resume on similar timescales (Figure~\ref{fig:ztf_season_skews}), necessitating the disk be replenished $\sim\,1-2$\,yr if indeed that is the maximum disk lifetime. It is challenging to replenish disks at such a rapid rate; among the most expeditious processes yet demonstrated is sesquinary catastrophe, still necessitating likely at least $\sim10^2$\,yr for the orbit at \obj\ \citep{Veras2025}. However, secular chaos have been demonstrated to be able to steadily inject planetary material into the Roche limit of white dwarfs for it to be accreted \citep{OConnor2022}, especially for young white dwarfs like \obj ($\tau_{\rm cool} \approx 110\,$Myr, as estimated using the {\tt WD\_models}\footnote{\url{https://github.com/SihaoCheng/WD_models}} implementation of the \citealt{Bedard2020} cooling models). Alternatively, the debris transiting \obj could have been delivered by a rotationally disrupted asteroid orbiting exterior to the nominal rubble-pile tidal disruption distance \citep{Veras2020}. The still-disrupting, intact asteroid could impart small perturbations onto the debris disk, occasionally nudging it out of and back into alignment with our line of sight \citep[e.g.,][]{Nesvold2016}\footnote{This picture would require the third body to be more massive than the disk, but not so massive as to blast the disk away from the star.}.  Interactions by an unseen perturber were invoked to explain the orbital resonances in the transiting debris at WD\,1054$-$226 \citep{Farihi2022}. The numerous potential harmonics that we report in Section~\ref{sec:transit_harmonics} could similarly lend credence to the possibility of interactions with an external body. Finally, general relativistic precession has been proposed in the gas disk at WD\,1145+017 \citep{Cauley2018} to occur on a timescale of about 5\,yr. It remains unclear how such precession is coupled (if at all) to the dusty components of the disk and if its effects propagate into the transits \citep{Miranda2018}.

\subsection{Colorless, Gray Transits}\label{sec:color}

The transit depths from our near-simultaneous $g$+$r$ and $g$+$i$ CHIMERA photometry are consistent with each other. In Figure~\ref{fig:color_dependence} we interpolate between the nearest blue and red points for each CHIMERA night from 2023~May, 2023~June, and 2023~August (only the 2023~May~21 light curve includes $i$-band observations) to compare the measured relative magnitudes. Our interpolation does not change the time stamps by more than 2.5\,s. The results are consistent with colorless transits, following a 1:1 trend. As done by \citet{Farihi2022}, we test this trend in the bottom panels of Figure~\ref{fig:color_dependence} by fitting lines using the orthogonal distance regression (ODR) method as implemented by {\tt scipy} \citep{2020SciPy-NMeth} to determine the extinction coefficients between the blue and red channels. The ODR technique is necessary because the $g$-band measurements and their uncertainties occupy both axes; a linear least squares fit would be skewed by these correlated variables. The best-fit slopes are $(g-r)/g = 0.002 \pm 0.004$ and $(g-i)/g = -0.010 \pm 0.012$, demonstrating the transit depths are consistent with being colorless, or gray. The residual variances\footnote{The residual variance, sometimes expressed as $\hat{\sigma}^2$, is a goodness of fit metric that is analogous to the reduced $\chi^2$ for a linear least squares fit. Formally, $\hat{\sigma}^2$ quantifies the deviation left unexplained by the linear regression.} of these fits are identical to those of fixed horizontal lines: for $(g-r)/g$ the residual variances for the best-fit and horizontal lines are 0.63 and 0.63, respectively, and for $(g-i)/g$: 0.75 and 0.75. These extinction coefficients are an order of magnitude less than what is found for interstellar dust ($((g-r)/g)_{\rm ISM}=0.287$; \citealt{Wang2019}; \citealt{Farihi2022}). This lack of color indicates a dearth of small dust particles in the transiting effluents at \obj, just like at WD\,1145+017 \citep{Alonso2016,Zhou2016,Croll2017,Xu2018a} and WD\,1054$-$226 \citep{Farihi2022}.

\begin{figure*}[t!]
    \epsscale{1.16}
    \plotone{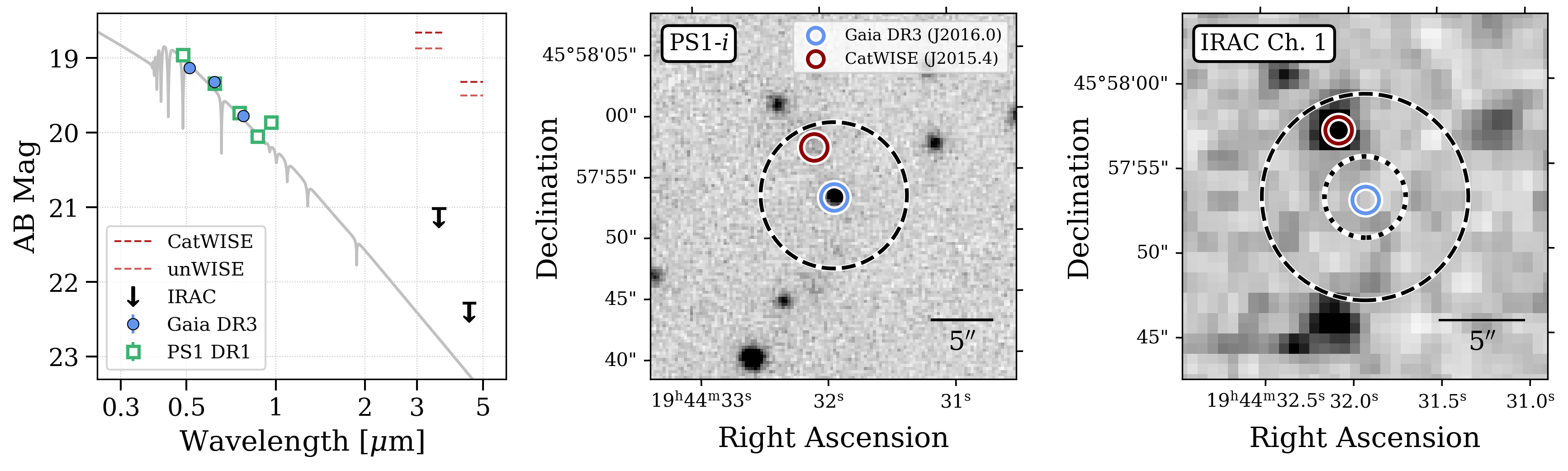}
    \caption{Left: SED of \obj including photometry from Gaia DR3, Pan-STARRS DR1, CatWISE, unWISE, and Warm Spitzer. We superimpose these photometry above a synthetic spectrum for a DA white dwarf with $T_{\rm eff} = 21{,}000\,$K and $\log(g)=8.25$ \citep{Koester2010}. The error bars on the Gaia and PS1 photometry are smaller than the marker sizes at these scales ($\sigma_m \lesssim 0.1\,$mag). Horizontal bars are spurious detections and downward-pointing arrows indicate upper limits. We are unable to currently place reliable constraints on the presence of an infrared excess at \obj. The middle (PS1 $i$-band) and right (Spitzer IRAC Channel~1) images elucidate the unreliability of these near-infrared photometry. The 6.1\arcsec-radius aperture of WISE is shown as a black dashed circle, centered on the Gaia-measured centroid of \obj (green circle) and enclosing the centroid of the nearest source to \obj as measured by CatWISE (maroon circle). The CatWISE and unWISE centroids are coincident with a bright field source as imaged by Spitzer. The smaller 2.4\arcsec radius we use for our Spitzer photometry is shown as the black dotted circle in the rightmost panel, illustrating how the pixels at the Gaia-measured centroid of \obj are indistinguishable from empty sky.
    }  \label{fig:SED}
\end{figure*}

We are presently unable to impose firm constraints on the particle size distribution of the transiting dusty effluents. \citet{Xu2018a}, when investigating the colorless transits at WD\,1145+017, demonstrate that dust grains with characteristic sizes $s$ that obey $X \equiv \frac{2 \pi s}{\lambda} \gtrsim 2$ extinguish starlight at equal efficiencies, yielding gray transits. Considering we observe consistent extinction from $0.4-0.8\,\mu$m, we can thus only tenuously estimate $s \gtrsim 0.2\,\mu$m. Dust grains in the local ISM tend to be $s \sim 0.2\,\mu$m \citep[e.g.,][]{Mathis1977,Kim1995}. Simultaneous optical and near-infrared photometry spanning a transit is needed to refine this limit into a reliable constraint.

Colorless transits are theoretically achievable even with the presence of some small dust grains. Modeling by \citet{Izquierdo2018} reproduces colorless transits at WD\,1145+017 while allowing for small particles within an edge-on, geometrically thin, optically thick disk. As disks veer slightly edge-off, reddening substantially grows, favoring a low {\AA}ngstr{\"o}m exponent ($\alpha \lesssim 0.5$) to satisfy colorlessness \citep{Bhattacharjee2025b}, all without expelling small dust grains from the disk (see Appendix~A of \citealt{Bhattacharjee2025b}).

\subsection{No Detected Infrared Excess}

We lack reliable constraints on the presence of an infrared excess at \obj. There is no publicly archived $JHK$ photometry that is readily accessible, attributable to the star's faintness. We find entries for \obj in the CatWISE \citep{Marocco2021} and unWISE  \citep{Schlafly2019} catalogs, both of which suggest a $>$5$\sigma$ excess of 3.4-$\mu$m emission relative to the model white dwarf spectrum (Figure~\ref{fig:SED}). We caution these are low-fidelity, likely spurious photometry that should not be treated as detections. While there is no Figure of Merit available to vet source confusion (\obj is not in the {\tt gaiadr3.allwise\_neighbourhood} table of \citealt{Marrese2022}; see \citealt{Dennihy2020a}), we find the unWISE and CatWISE centroids are separated from the Gaia-measured centroid by 4.2\arcsec\ and 4.4\arcsec, respectively, both coincident with a bright field source and still comfortably within the $\approx$6.1\arcsec\ WISE PSF at $W1$ (Figure~\ref{fig:SED}). There is public Warm Spitzer IRAC imaging of \obj at 3.6\,$\mu$m and 4.5\,$\mu$m (Spitzer GO \#10067; \citealt{Werner2021}) circa 2014~October, which we obtain via the Spitzer Heritage Archive at IRSA. Both images were integrated to an effective exposure time of 10.4\,s per pixel. We do simple circular aperture photometry with {\tt photutils} \citep{larry_bradley_2024_13989456} on these images using circular 2.4\arcsec-radius apertures (2\,pixels, equivalently) for both the Channel 1 and Channel 2 images with background annuli extending from $r_{\rm in} = 12$\,pix to $r_{\rm out} = 20$\,, mirroring the photometric extraction by \citet{Werner2021}. We raise the caveat that numerous of the flux-calibrated pixel values within our 2.4\arcsec aperture and background annuli are negative. We incorporate the Spitzer error images into the {\tt photutils} photometry extraction to estimate uncertainties. We calculate upper limits on the Channel 1 and Channel 2 AB magnitudes of $m_{\rm Ch1} \geq 21.01$\,mag and $m_{\rm Ch2} \geq 22.28$\,mag, utilizing the published photometric conversions provided by IRSA and the IRAC Instrument Handbook\footnote{\url{https://irsa.ipac.caltech.edu/data/SPITZER/docs/irac/iracinstrumenthandbook/}} (see also \citealt{Reach2005}). These upper limits are consistent with blank regions of the sky, at R.A.\,$= 296.129167$\,deg and decl.\,$= 45.970000$\,deg, where we calculate $m_{\rm Ch1} = 22.14 \pm 0.60$\,mag and $m_{\rm Ch2} = 22.75 \pm 0.53$\,mag. We are unable to stringently constrain the presence of excess near-infrared emission at \obj to within the limits of these observations.

We can estimate an upper limit on the mass of the dusty component of the disk \obj is accreting from the upper limit of the flux we measure at 3.6\,$\mu$m from IRAC. We construct the following toy model for a disk using best-determined parameters of the dust disk at G29$-$38 by \citet{Ballering2022}: a flat, cylindrical ring with $r_{\rm in} = a - 0.5\,{\rm R_{WD}}$, $r_{\rm out} = a + 0.5\,{\rm R_{WD}}$, and a height of $H = 2.2 R_{\rm WD}$, with a density of $\rho = 1.5 \cdot 10^{-12}$\,g\,cm$^{-3}$ (provided a total disk mass of $M = 10^{18}$\,g; c.f., \citealt{LeBourdais2024}). Here, $a=1.3618$\,\rsun, the semi-major axis found for a circular orbit at 4.9704\,hr, and $R_{\rm  WD} = 0.0102$\,\rsun (see Section~\ref{sec:orbit}), meaning the dust would be heated to $T_{\rm eq} = 1395$\,K (Section~\ref{sec:orbit}). To estimate the mass of this disk from the IRAC 3.6-$\mu$m photometry we evaluate $M_{\rm dust} = F_{\nu} d^2 / (B_{\nu}(T_{\rm eq}) \kappa_{\nu})$, where $d = 447^{+44}_{-40}\,$pc \citep{BailerJones2021} and $\kappa_{\nu} = 3 Q_{\nu} / (4s\rho)$ \citep{Zuckerman2001}. For $s=0.2\,\mu$m silicates, $Q_{\nu} \approx 0.03$ at $\lambda = 3.6\,\mu$m \citep{Draine&Lee1984}, yielding $M_{\rm dust} \leq 4 \cdot10^{15}$\,g. This simple estimation applies only for optically thin dust. We currently lack viable constraints to the optical depth of the debris and its geometry.

Among the transiting white dwarfs with hours-long orbital periods, neither ZTF\,J0328$-$1219 nor WD\,1054$-$226 show near-infrared excesses, unlike WD\,1145+017 and SBSS\,1232+563. This could just be a reflection of the observation that metal-rich white dwarfs are rarely accompanied by significant infrared emission from the dusty disk \citep{Rocchetto2015,Wilson2019}; the transiting systems suffer an additional inclination bias towards more edge-on viewing geometry that may make it harder to detect infrared excess \citep[see, e.g.,][]{Bhattacharjee2025b}. Deeper infrared imaging from the ground at $JHK$ or JWST observations are necessary to apply more viable constraints on emission from a close-in, dusty debris disk at \obj.

\begin{deluxetable*}{lccccccccc}[!ht]
\tablenum{2}
\tablecaption{Select Properties of the White Dwarfs Showing Periodic Transiting Debris \label{tab:transit_systems}}
\tabletypesize{\footnotesize}
\tablehead{
    \colhead{White Dwarf}    & \colhead{Orbital Period}  &
    \colhead{IR Excess}   & \colhead{CS Gas} & 
    \colhead{Transit Depth} & \colhead{Spec. Type} &
    \colhead{$T_{\rm eff}$ (K)} & \colhead{Mass (\msun)} &
    \colhead{$\tau_{\rm cool}$}     & \colhead{$\tau_{\rm variations}$} 
}
\startdata
\obj$^{\rm a}$  &  4.9704\,hr &  N  &  N  & $\approx 40\%$ & DAZ & 20${,}$790 & 0.79 & 110\,Myr & Single orbit  \\
\hline
WD\,1145+017$^{\rm b}$  &  4.49\,hr &  Y  &  Y & $\approx 40\%$* & DBZA & 15${,}$500$^{\rm g}$ & 0.66$^{\rm h}$  & 180\,Myr & Single orbit \\
ZTF\,J0328$-$1219$^{\rm c}$  &  9.937\,hr &  N  & Y & $\approx 10\%$ & DZ & 7630 & 0.73 & 1.84\,Gyr & Weeks  \\
SBSS\,1232+563$^{\rm d}$  &  14.842\,hr &  Y  &  N & $\approx 5\%$ & DBZA & 11${,}$790$^{\rm i}$ & 0.77$^{\rm i}$  & 640\,Myr$^{\rm i}$  & Unknown \\
WD\,1054$-$226$^{\rm e}$  &  25.02\,hr &  N  &  N & $\approx 5\%$ & DAZ & 7910 & 0.62 & 1.3\,Gyr & Weeks\\
ZTF\,J0139+5245$^{\rm f}$  &  107.2\,d &  N  &  ? & $\approx 40\%$ & DAZ & 10${,}$530 & 0.52 & 460\,Myr$^{\rm j}$ & Single orbit?
\enddata
\tablecomments{Discovery papers: a -- This work; b -- \citet{Vanderburg2015}; c -- \citet{Vanderbosch2021}; d -- \citet{Hermes2025}; e -- \citet{Farihi2022}; f -- \citet{Vanderbosch2020}. We report only the fundamental orbital periods, and not any additional ``B" periods or periods of drifting fragments; at least WD\,1145+017, ZTF\,J0328$-$1219, and WD\,1054$-$226 show additional independent periods. Additional references: g -- \citet{LeBourdais2024}; h -- \citet{Fortin-Archambault2020}; i -- \citet{Coutu2019}; j -- \citet{Bedard2020} via {\tt WD\_models}. ZTF\,J0139+5245 might show circumstellar gas, hence its classification as ?. $\tau_{\rm variations}$ refers to the timescale upon which changes in transit morphologies are observed. This timescale at ZTF\,J0139+5245 is questioned because its period has shown to be unstable over a single orbit (see Figure~5 of \citealt{Vanderbosch2020}). We denote the asterisk on the transit depths of WD\,1145+017 because the debris subtends a larger fraction of the stellar cross-sectional area compared to the gas disk, driving a shallow in-transit bluing that is especially noticeable in the far-UV \citep{Xu2019a}. The transits at each system appear gray.}
\end{deluxetable*}
\vspace{-0.8cm}

\subsection{Orbital Constraints}\label{sec:orbit}

Similar to WD\,1145+017, we estimate the debris transiting \obj\ is orbiting just outside the tidal disruption radius of the star. Using the framework from \citet{Veras2014,Veras2017} and assuming $M_{\rm WD} = 0.79\,$M$_{\odot}$ \citep{Bhattacharjee2025}, we calculate that synchronously rotating and non-rotating bodies (see Table~1 of \citealt{Veras2017}) with asteroid-like densities ($\rho=3$\,g\,cm$^{-3}$) would tidally disrupt upon approaching separations of 0.90\,R$_{\odot}$ and 1.03\,R$_{\odot}$, respectively, which equate to circular orbits at 2.39 and 2.92 hours (these inflate to 3.30 and 4.02 hours for an asteroid with $\rho=2$\,g\,cm$^{-3}$). We estimate this toy asteroid would orbit at a eccentricity no smaller than $e_{\rm min} \geq 0.33$ by emulating the framework of \citet{Farihi2022} (see their Figure~11). 

The projected size of the rubble following this orbit must be large relative to the cross-section of the white dwarf. A single, intact, opaque spherical object that casts a 40\% transit depth would require a radius of about 4500\,km (about $0.63\,{\rm R_ {WD}}$), estimated by using the formulae for white dwarf radii developed by \citet{Nauenberg1972,Verbunt1988,Veras2014}. The fact that there are multiple inflection points within the longest-duration and deepest transits suggests there is probably a superposition of smaller objects clumped together that are occulting starlight, which appears to be a plausible model for WD\,1145+017 (\citealt{Gaensicke2016}; \citealt{Veras2017}; Figure~7 of \citealt{Izquierdo2018}). Our earlier calculations of the tidal disruption radii at \obj\ did not account for the internal tensile strength of the parent object. Following the initial disruption event, fragmented particles can withstand tidal disruption at the nominal Roche limit \citep{McDonald2021}; collisions and erosion are needed to grind them down finer in the absence of inward migration. The transits we observe could be the clumps of such fragments, which on their measured orbit would likely have yet to cross the sublimation radius of the star \citep{Veras2023}, although it seems some material must be sublimated to match the observed evidence for circumstellar gas. The best-determined orbital period of 4.9704\,hr implies a semi-major axis of $a = 1.3618$\,R$_{\odot}$ by Kepler's Third Law. The radiation field of a $T_{\rm eff} = 20{,}790$\,K blackbody would heat material orbiting circularly along this semi-major axis to an equilibrium temperature of $T_{\rm eff} = 1395$\,K, almost sufficiently hot to sublimate calcium \citep{2011EP&S...63.1067K,2024A&A...692A..25F}. Iron would likely remain solid at this temperature \citep{Rafikov&Garmilla2012,vanLieshout2014}, needing to overcome a sublimation radius of $r_{\rm sub, Fe} = 0.84$\,\rsun\ for this system \citep{Veras2022}.

\section{Conclusions \& Outlook}\label{sec:conclusion}

We have obtained more than 123 hours of high-speed time-series photometry of the metal-polluted (DAZ) white dwarf \obj\ across a baseline spanning 2022~March through 2025~July. Our dense follow-up taken during 2023~May through 2023~November shows repeating transits every $4.9704 \pm 0.0019$\,hr, making \obj the second known case of transiting planetary debris near the Roche limit of a white dwarf. Transits vanished from 2024~May through 2024~October, but appear to recover in 2025~March -- repeating again at 4.9704\,hr, marking the second observation of a cessation and perhaps the first resumption of periodic transits at a white dwarf remnant planetary system. This fluctuation in transit activity appears corroborated by the ZTF light curve, showing a strongly negative skew during an era of detected transits in mid-2023 and near-zero skew in 2024 when transits were absent, before again veering to negative skew and observed transits in early 2025. The cause of the variable transit activity is presently unknown.

Time-resolved spectroscopy concurrent with our time-series photometry on 2023~June~20 shows the Ca-K absorption line does not vary significantly from in-transit to out-of-transit, possibly suggesting a lack of circumstellar gas within the transiting effluents. At the same time, the presence of some of the observed metal lines and their equivalent widths cannot be reproduced by our synthetic spectrum, likely demonstrating there is circumstellar gas orbiting \obj. We do not observe evidence for small dust particles within the transiting debris, evidenced by the lack of an in-transit color term. Nor do we detect an infrared excess in public near-infrared Warm Spitzer IRAC photometry. We compare the properties of \obj to the five other known cases of periodic transits from planetary debris at white dwarfs in Table~\ref{tab:transit_systems}. We see an interesting possible correlation, where transiting white dwarfs may collectively show a mean mass in excess to field white dwarfs ($\langle M_{\rm WD} \rangle \approx 0.6$\,\msun; see, e.g., \citealt{OBrien2024}). This trend could in part be explained by transit occultations creating apparently underluminous photometry, which could be leveraged to discover additional candidate white dwarfs with transiting debris.

Our measured orbital period of 4.9704\,hr is predicted to be rare for transiting tidally disrupted debris at white dwarfs (see Figure~11 of \citealt{Li2025a}). In this picture, the transits we observe at \obj could be a window into highly evolved debris, whose orbit has circularized and migrated inwards over myriad cycles of tidal disruption following its initial injection into the Roche sphere on a highly eccentric orbit \citep[e.g.,][]{Brouwers2022,Li2025a,Li2025b}. Similar to the star's analog of WD\,1145+017, the implied rapid dynamical and morphological evolution of the transits establish \obj as an exciting laboratory for further constraining how white dwarf debris disks can be generated via tidal disruption, as well as applying further constraints to transit detection rates and evolutionary models of these systems. Most compelling is the prospect of testing models that can reproduce the observed cessation and resumption of transits.  

We enthusiastically await the imminent stream of photometry from upcoming time-domain surveys that will greatly benefit the study of transiting white dwarf remnant planetary systems, namely by the Rubin Observatory \citep{2019ApJ...873..111I} and the Roman Space Telescope \citep{2019arXiv190205569A}. Infrared transit photometry from Roman could provide the requisite wavelength coverage to begin to place constraints on the particle size distribution of the debris at transiting debris systems like \obj via Mie scattering theory \citep[c.f.,][]{Xu2018a}. \obj may be just within the projected threshold for $\sim$1\%-level photometry at the F146W filter within single exposures by Roman (\citealt{Tamburo2023}; see Figure~\ref{fig:SED}).

\section{Acknowledgments}

We first extend our gratitude to our anonymous referee, whose careful review and recommendations enhanced this manuscript. In fruitful conversations and correspondence with Tim Cunningham, Jay Farihi, Jim Fuller, Philip Muirhead, Saul Rappaport, Siyi Xu \begin{CJK*}{UTF8}{gbsn}(许\CJKfamily{bsmi}偲\CJKfamily{gbsn}艺\end{CJK*}), and Nadia Zakamska, we found guidance that improved our interpretation of these results. We are deeply grateful for the observing support by John Kuehne at McDonald Observatory and Colt Pauley at the Perkins Telescope Observatory. This material is based upon work supported by the National Aeronautics and Space Administration under Grant No. 80NSSC23K1068 issued through the Science Mission Directorate. J.\,A.\,G. is supported by the National Science Foundation Graduate Research Fellowship Program under Grant No. 2234657. 

This worked is based on observations obtained with the Samuel Oschin Telescope 48-inch and the 60-inch Telescope at the Palomar Observatory as part of the Zwicky Transient Facility project. ZTF is supported by the National Science Foundation under Grants No. AST-1440341 and AST-2034437 and a collaboration including current partners Caltech, IPAC, the Oskar Klein Center at Stockholm University, the University of Maryland, University of California, Berkeley , the University of Wisconsin at Milwaukee, University of Warwick, Ruhr University, Cornell University, Northwestern University and Drexel University. Operations are conducted by COO, IPAC, and UW. 

This work has made use of data from the European Space Agency (ESA) mission {\it Gaia} (\url{https://www.cosmos.esa.int/gaia}), processed by the {\it Gaia} Data Processing and Analysis Consortium (DPAC, \url{https://www.cosmos.esa.int/web/gaia/dpac/consortium}). Funding for the DPAC has been provided by national institutions, in particular the institutions participating in the {\it Gaia} Multilateral Agreement.

This publication also makes use of data products from NEOWISE, which is a project of the Jet Propulsion Laboratory/California Institute of Technology, funded by the Planetary Science Division of the National Aeronautics and Space Administration.
 
This work is based in part on observations made with the Spitzer Space Telescope, which was operated by the Jet Propulsion Laboratory, California Institute of Technology under a contract with NASA.

The Pan-STARRS1 Surveys (PS1) and the PS1 public science archive have been made possible through contributions by the Institute for Astronomy, the University of Hawaii, the Pan-STARRS Project Office, the Max-Planck Society and its participating institutes, the Max Planck Institute for Astronomy, Heidelberg and the Max Planck Institute for Extraterrestrial Physics, Garching, The Johns Hopkins University, Durham University, the University of Edinburgh, the Queen's University Belfast, the Harvard-Smithsonian Center for Astrophysics, the Las Cumbres Observatory Global Telescope Network Incorporated, the National Central University of Taiwan, the Space Telescope Science Institute, the National Aeronautics and Space Administration under Grant No. NNX08AR22G issued through the Planetary Science Division of the NASA Science Mission Directorate, the National Science Foundation Grant No. AST-1238877, the University of Maryland, Eotvos Lorand University (ELTE), the Los Alamos National Laboratory, and the Gordon and Betty Moore Foundation.

This research relied upon the SIMBAD and VizieR databases operated by CDS (Strasbourg, France) and the bibliographic resources of The SAO Astrophysics Data System.


\vspace{5mm}
\facilities{Zwicky Transient Facility, Palomar Observatory: Hale 200inch (CHIMERA, DBSP), McDonald Observatory: Struve 2.1m (ProEM), Perkins Telescope Observatory (PRISM), Lowell Discovery Telescope (LMI), Keck:I (LRIS), Gaia, PS1, Spitzer, WISE}

\software{Astropy \citep{astropy:2013, astropy:2018, astropy:2022}, astroquery \citep{2019AJ....157...98G}, ccdproc \citep{matt_craig_2017_1069648}, cuvarbase \citep{cuvarbase2022}, extinction \citep{exctinction}, hipercam \citep{Dhillon2021}, lmfit \citep{LMFIT_2014}, matplotlib \citep{Hunter:2007}, numpy \citep{harris2020array}, pandas \citep{mckinney-proc-scipy-2010,the_pandas_dev_team_2024_13819579}, phot2lc \citep{phot2lc}, photutils \citep{larry_bradley_2024_13989456}, Pyriod \citep{Pyriod2022}, scipy \citep{2020SciPy-NMeth}}

\appendix

\section{Observing Log for Time-Series Photometry}\label{sec:ts_phot_appendix}

We provide in Table~\ref{tab:ts_phot} a record of our follow-up time-series photometry observations in chronological order.

\newpage

\begin{deluxetable*}{lccccccc}[htbp!]
\tablenum{3}
\tablecaption{Time Series Photometry Observations \label{tab:ts_phot}}
\tabletypesize{\footnotesize}
\tablehead{
    \colhead{UT Date}    & \colhead{Facility}  &
    \colhead{Duration}   & \colhead{$t_{\mathrm{exp}}$} & 
    \colhead{Filter}     & \colhead{Average Seeing} & 
    \colhead{$\Delta F$} & \colhead{Periodograms}  \\ [-0.2cm]
    \colhead{}           & \colhead{}      &
    \colhead{(hr)}       & \colhead{(s)}   &  \colhead{} &
    \colhead{(arcsec)}   & \colhead{\%}    &  \colhead{}
}
\startdata
2022 Mar 23  &  P200/CHIMERA      &  1.26    &  10  &  $g$      &  2.90  & 13.4  &     \\
             &  P200/CHIMERA      &  1.26    &  10  &  $r$      &  2.61  & 14.4  &     \\
2022 Apr 29  &  McD/ProEM         &  2.58    &  15  &  $BG40$   &  1.58  & 32.8  &     \\
2022 Sep 29  &  P200/CHIMERA      &  1.54    &  10  &  $g$      &  1.20  & 4.6   &     \\
             &  P200/CHIMERA      &  1.54    &  10  &  $r$      &  1.12  & 11.6  &     \\
2023 May 21  &  P200/CHIMERA      &  3.21    &  10  &  $g$      &  1.13  & 34.5  &   Y \\
             &  P200/CHIMERA      &  3.21    &  10  &  $r$      &  0.96  & 34.3  &     \\
2023 May 25  &  P200/CHIMERA      &  5.63    &  10  &  $g$      &  1.64  & 35.7  &   Y \\
             &  P200/CHIMERA      &  5.67    &  10  &  $i$      &  1.34  & 41.2  &     \\
2023 Jun 19  &  P200/CHIMERA      &  7.42    &  10  &  $g$      &  1.34  & 38.3  &   Y \\
             &  P200/CHIMERA      &  7.43    &  10  &  $r$      &  1.13  & 40.8  &     \\
             &  McD/ProEM         &  1.64    &  10  &  $BG40$   &  0.80  & 30.7  &     \\
2023 Jun 20  &  P200/CHIMERA      &  7.67    &  10  &  $g$      &  1.38  & 43.8  &   Y \\
             &  P200/CHIMERA      &  7.68    &  10  &  $r$      &  1.17  & 41.5  &     \\
             &  McD/ProEM         &  2.56    &  15  &  $BG40$   &  1.08  & 42.7  &     \\
2023 Jun 29  &  McD/ProEM         &  5.61    &  46  &  $BG40$   &  1.85  & 36.9  &   Y \\
2023 Aug 12  &  P200/CHIMERA      &  6.05    &  10  &  $g$      &  1.10  & 35.7  &   Y \\
             &  P200/CHIMERA      &  6.05    &  10  &  $r$      &  0.92  & 41.8  &     \\
2023 Sep 20  &  McD/ProEM         &  5.88    &  20  &  $BG40$   &  1.04  & 31.4  &   Y \\
2023 Sep 21  &  McD/ProEM         &  0.39    &  15  &  $BG40$   &  1.12  & 29.1  &   Y \\
2023 Sep 22  &  McD/ProEM         &  2.85    &  15  &  $BG40$   &  1.06  & 40.4  &   Y \\
2023 Oct 15  &  PTO/PRISM         &  4.00    &  30  &  $BG40$   &  2.61  & 51.9  &   Y \\
2023 Oct 16  &  PTO/PRISM         &  3.22    &  40  &  $BG40$   &  2.66  & 35.4  &   Y \\
2023 Nov 14  &  PTO/PRISM         &  1.48    &  35  &  $BG40$   &  2.09  & 22.7  &   Y \\
2023 Nov 15  &  PTO/PRISM         &  2.19    &  35  &  $BG40$   &  2.13  & 27.6  &   Y \\
2024 May 14  &  P200/CHIMERA      &  3.66    &  10  &  $g$      &  0.89  & 3.5   &     \\
             &  P200/CHIMERA      &  3.66    &  10  &  $r$      &  0.67  & 3.1   &     \\
2024 Jul 13  &  P200/CHIMERA      &  4.75    &  10  &  $g$      &  0.93  & 3.6   &     \\
             &  P200/CHIMERA      &  4.75    &  10  &  $r$      &  0.82  & 5.9   &     \\
2024 Jul 14  &  P200/CHIMERA      &  4.67    &  10  &  $g$      &  1.08  & 5.7   &     \\
             &  P200/CHIMERA      &  4.67    &  10  &  $r$      &  1.38  & 7.4   &     \\
2024 Aug 03  &  PTO/PRISM         &  2.02    &  30  &  $BG40$   &  2.72  & 11.2  &     \\
2024 Aug 06  &  PTO/PRISM         &  1.15    &  30  &  $BG40$   &  2.00  & 6.3   &     \\
             &  PTO/PRISM         &  1.42    &  35  &  $BG40$   &  2.10  & 9.4   &     \\
2024 Aug 10  &  PTO/PRISM         &  0.87    &  40  &  $BG40$   &  2.14  & 5.5   &     \\
2024 Aug 25  &  LDT/LMI           &  1.82    &  15  &  $g$      &  1.14  & 3.1   &     \\
2024 Aug 26  &  PTO/PRISM         &  5.23    &  40  &  $BG40$   &  2.42  & 8.2   &     \\
2024 Aug 27  &  PTO/PRISM         &  5.49    &  40  &  $BG40$   &  2.57  & 6.7   &     \\
2024 Oct 31  &  PTO/PRISM         &  3.17    &  60  &  $BG40$   &  2.91  & 8.7   &     \\
2024 Nov 01  &  PTO/PRISM         &  2.21    &  45  &  $BG40$   &  2.62  & 14.2  &     \\
             &  PTO/PRISM         &  0.90    &  60  &  $BG40$   &  2.89  & 6.8   &     \\
2025 Mar 24  &  McD/ProEM         &  1.70    &  30  &  $BG40$   &  1.65  & 53.1  &     \\
2025 Mar 25  &  McD/ProEM         &  1.82    &  30  &  $BG40$   &  1.83  & 55.2  &     \\
2025 May 27  &  McD/ProEM         &  3.38    &  20  &  $BG40$   &  1.91  & 56.9  &     \\
             &  McD/ProEM         &  0.48    &  30  &  $BG40$   &  2.50  & 24.9  &     \\
2025 May 29  &  McD/ProEM         &  4.63    &  30  &  $BG40$   &  1.18  & 45.5  &     \\
2025 Jul 30  &  McD/ProEM         &  5.11    &  30  &  $BG40$   &  0.95  & 42.4  &     \\
2025 Jul 31  &  McD/ProEM         &  6.99    &  30  &  $BG40$   &  0.89  & 42.5  &     \\
\enddata
\tablecomments{In the Facility column, P200 is the Palomar Hale 200-in telescope, McD is the McDonald 2.1-m Otto Struve telescope, PTO is the Perkins 1.8-m telescope, and LDT is the 4.3-m Lowell Discovery Telescope. Our seeing estimates are derived from our reductions using {\tt hipercam}. $\Delta F$ scores the difference between the median of the flux points within the upper quintile and minimum flux point of each light curve after binning to 1-minute exposures, effectively a proxy for the greatest transit depth observed at that epoch. The Periodogram column indicates which light curves were used to calculate periodograms and refine the orbital period in Section~\ref{sec:period}.}
\end{deluxetable*}

\onecolumngrid

\section{Observing Log for Spectroscopy}\label{sec:spec_appendix}

We provide in Table~\ref{tab:spec_obs} a record of our follow-up spectroscopic observations in chronological order.

\newpage

\begin{deluxetable*}{cccccccccccD}[htbp!]
\tablenum{4}
\tablecaption{Summary of Spectroscopic Observations \label{tab:spec_obs}}
\tabletypesize{\footnotesize}
\tablehead{
    \colhead{Date}                       & \colhead{Time\textbf{}}          & \colhead{Facility} &  \colhead{$t_{\mathrm{exp}}$}         & \colhead{Slit Width}             & \colhead{Seeing}   &
    \colhead{Airmass}                    & \colhead{$\lambda_{\mathrm{s}}$} & \colhead{$R$}      & 
    \colhead{$\mathrm{W_{CaK}}$}         & \colhead{S/N}                    & \colhead{$\langle f\rangle$} \\ [-0.2cm]
    \colhead{(WD\,J)}                    & \colhead{(UTC)}                  & \colhead{}   &
    \colhead{(s)}                        & \colhead{($''$)}                 & \colhead{($''$)}  &
    \colhead{}                           & \colhead{(\AA)}                  & \colhead{($\lambda/\Delta\lambda$)} & 
    \colhead{(\AA)}                      & \colhead{}                       & \colhead{(\%)}
}
\startdata
2022 May 06  & 11:03:52     &  P200${+}$DBSP  &  2$\times$600 &  1.5  & 1.2 & 1.07  & 5627  &  1060(B)/960(R)   &  $0.79\pm0.29$  & 8.7 &             \\
2022 Jul 04  & 10:08:18     &  Keck${+}$LRIS  &  600          &  1.0  & 1.3 & 1.15  & 5643  &  1547(B)/912(R)   &  $0.65\pm0.33$  & 11.5 &            \\
$\cdots$     & 10:18:58     &  Keck${+}$LRIS  &  600          &  1.0  & 1.3 & 1.14  & 5643  &  1549(B)/937(R)   &  $1.34\pm0.31$  & 11.7 &            \\
$\cdots$     & 10:29:35     &  Keck${+}$LRIS  &  600          &  1.0  & 1.2 & 1.13  & 5643  &  1544(B)/943(R)   &  $0.61\pm0.26$  & 12.8 &            \\
2022 Nov 21  & 02:37:32     &  P200${+}$DBSP  &  3$\times$900 &  1.5  & 1.4 & 1.32  & 5773  &  1060(B)/960(R)   &  $0.99\pm0.29$  & 8.8 &          \\
2023 Jun 20  & 10:48:44     &  Keck${+}$LRIS  &  600          &  1.0  & 1.3 & 1.17  & 5643  &  1471(B)/1550(R)  &  $0.48\pm0.29$  & 9.7 & $-27.4{\pm}0.4$  \\
$\cdots$     & 10:59:21     &  Keck${+}$LRIS  &  600          &  1.0  & 1.1 & 1.15  & 5643  &  1478(B)/1557(R)  &  $0.65\pm0.24$ & 10.4 & $-23.0{\pm}0.4$  \\
$\cdots$     & 11:09:58     &  Keck${+}$LRIS  &  600          &  1.0  & 1.0 & 1.14  & 5643  &  1473(B)/1542(R)  &  $0.48\pm0.22$  & 11.0 & $-15.5{\pm}0.4$ \\
$\cdots$     & 11:20:35     &  Keck${+}$LRIS  &  600          &  1.0  & 1.0 & 1.14  & 5644  &  1484(B)/1542(R)  &  $0.40\pm0.20$  & 11.6 &  $0.7{\pm}0.5$    \\
$\cdots$     & 11:31:12     &  Keck${+}$LRIS  &  600          &  1.0  & 0.9 & 1.13  & 5644  &  1483(B)/1547(R)  &  $0.83\pm0.19$  & 11.7 & $0.8{\pm}0.5$   \\
\enddata
\tablecomments{$\lambda_{\mathrm{s}}$ represents the wavelength at which the blue and red arms of the spectrographs are spliced together. Spectral resolutions for each arm are given in the $R$ column. S/N is the average signal-to-noise ratio per resolution element across the continua at $4600-4700$\,\AA. When concurrent time series photometry is available, $\langle f\rangle$ represents the average relative flux measured during the spectroscopic exposure.}
\end{deluxetable*}


\bibliography{ZTFJ1944+4557}{}

\begin{thebibliography}{}
\expandafter\ifx\csname natexlab\endcsname\relax\def\natexlab#1{#1}\fi
\providecommand{\url}[1]{\href{#1}{#1}}
\providecommand{\dodoi}[1]{doi:~\href{http://doi.org/#1}{\nolinkurl{#1}}}
\providecommand{\doeprint}[1]{\href{http://ascl.net/#1}{\nolinkurl{http://ascl.net/#1}}}
\providecommand{\doarXiv}[1]{\href{https://arxiv.org/abs/#1}{\nolinkurl{https://arxiv.org/abs/#1}}}

\bibitem[{{Akeson} {et~al.}(2019){Akeson}, {Armus}, {Bachelet}, {Bailey},
  {Bartusek}, {Bellini}, {Benford}, {Bennett}, {Bhattacharya}, {Bohlin},
  {Boyer}, {Bozza}, {Bryden}, {Calchi Novati}, {Carpenter}, {Casertano},
  {Choi}, {Content}, {Dayal}, {Dressler}, {Dor{\'e}}, {Fall}, {Fan}, {Fang},
  {Filippenko}, {Finkelstein}, {Foley}, {Furlanetto}, {Kalirai}, {Gaudi},
  {Gilbert}, {Girard}, {Grady}, {Greene}, {Guhathakurta}, {Heinrich},
  {Hemmati}, {Hendel}, {Henderson}, {Henning}, {Hirata}, {Ho}, {Huff},
  {Hutter}, {Jansen}, {Jha}, {Johnson}, {Jones}, {Kasdin}, {Kelly}, {Kirshner},
  {Koekemoer}, {Kruk}, {Lewis}, {Macintosh}, {Madau}, {Malhotra}, {Mandel},
  {Massara}, {Masters}, {McEnery}, {McQuinn}, {Melchior}, {Melton},
  {Mennesson}, {Peeples}, {Penny}, {Perlmutter}, {Pisani}, {Plazas}, {Poleski},
  {Postman}, {Ranc}, {Rauscher}, {Rest}, {Roberge}, {Robertson}, {Rodney},
  {Rhoads}, {Rhodes}, {Ryan}, {Sahu}, {Sand}, {Scolnic}, {Seth}, {Shvartzvald},
  {Siellez}, {Smith}, {Spergel}, {Stassun}, {Street}, {Strolger}, {Szalay},
  {Trauger}, {Troxel}, {Turnbull}, {van der Marel}, {von der Linden}, {Wang},
  {Weinberg}, {Williams}, {Windhorst}, {Wollack}, {Wu}, {Yee}, \&
  {Zimmerman}}]{2019arXiv190205569A}
{Akeson}, R., {Armus}, L., {Bachelet}, E., {et~al.} 2019, arXiv e-prints,
  arXiv:1902.05569, \dodoi{10.48550/arXiv.1902.05569}

\bibitem[{{Alonso} {et~al.}(2016){Alonso}, {Rappaport}, {Deeg}, \&
  {Palle}}]{Alonso2016}
{Alonso}, R., {Rappaport}, S., {Deeg}, H.~J., \& {Palle}, E. 2016, \aap, 589,
  L6, \dodoi{10.1051/0004-6361/201628511}

\bibitem[{{Astropy Collaboration} {et~al.}(2013){Astropy Collaboration},
  {Robitaille}, {Tollerud}, {Greenfield}, {Droettboom}, {Bray}, {Aldcroft},
  {Davis}, {Ginsburg}, {Price-Whelan}, {Kerzendorf}, {Conley}, {Crighton},
  {Barbary}, {Muna}, {Ferguson}, {Grollier}, {Parikh}, {Nair}, {Unther},
  {Deil}, {Woillez}, {Conseil}, {Kramer}, {Turner}, {Singer}, {Fox}, {Weaver},
  {Zabalza}, {Edwards}, {Azalee Bostroem}, {Burke}, {Casey}, {Crawford},
  {Dencheva}, {Ely}, {Jenness}, {Labrie}, {Lim}, {Pierfederici}, {Pontzen},
  {Ptak}, {Refsdal}, {Servillat}, \& {Streicher}}]{astropy:2013}
{Astropy Collaboration}, {Robitaille}, T.~P., {Tollerud}, E.~J., {et~al.} 2013,
  \aap, 558, A33, \dodoi{10.1051/0004-6361/201322068}

\bibitem[{{Astropy Collaboration} {et~al.}(2018){Astropy Collaboration},
  {Price-Whelan}, {Sip{\H{o}}cz}, {G{\"u}nther}, {Lim}, {Crawford}, {Conseil},
  {Shupe}, {Craig}, {Dencheva}, {Ginsburg}, {Vand erPlas}, {Bradley},
  {P{\'e}rez-Su{\'a}rez}, {de Val-Borro}, {Aldcroft}, {Cruz}, {Robitaille},
  {Tollerud}, {Ardelean}, {Babej}, {Bach}, {Bachetti}, {Bakanov}, {Bamford},
  {Barentsen}, {Barmby}, {Baumbach}, {Berry}, {Biscani}, {Boquien}, {Bostroem},
  {Bouma}, {Brammer}, {Bray}, {Breytenbach}, {Buddelmeijer}, {Burke},
  {Calderone}, {Cano Rodr{\'\i}guez}, {Cara}, {Cardoso}, {Cheedella}, {Copin},
  {Corrales}, {Crichton}, {D'Avella}, {Deil}, {Depagne}, {Dietrich}, {Donath},
  {Droettboom}, {Earl}, {Erben}, {Fabbro}, {Ferreira}, {Finethy}, {Fox},
  {Garrison}, {Gibbons}, {Goldstein}, {Gommers}, {Greco}, {Greenfield},
  {Groener}, {Grollier}, {Hagen}, {Hirst}, {Homeier}, {Horton}, {Hosseinzadeh},
  {Hu}, {Hunkeler}, {Ivezi{\'c}}, {Jain}, {Jenness}, {Kanarek}, {Kendrew},
  {Kern}, {Kerzendorf}, {Khvalko}, {King}, {Kirkby}, {Kulkarni}, {Kumar},
  {Lee}, {Lenz}, {Littlefair}, {Ma}, {Macleod}, {Mastropietro}, {McCully},
  {Montagnac}, {Morris}, {Mueller}, {Mumford}, {Muna}, {Murphy}, {Nelson},
  {Nguyen}, {Ninan}, {N{\"o}the}, {Ogaz}, {Oh}, {Parejko}, {Parley}, {Pascual},
  {Patil}, {Patil}, {Plunkett}, {Prochaska}, {Rastogi}, {Reddy Janga},
  {Sabater}, {Sakurikar}, {Seifert}, {Sherbert}, {Sherwood-Taylor}, {Shih},
  {Sick}, {Silbiger}, {Singanamalla}, {Singer}, {Sladen}, {Sooley},
  {Sornarajah}, {Streicher}, {Teuben}, {Thomas}, {Tremblay}, {Turner},
  {Terr{\'o}n}, {van Kerkwijk}, {de la Vega}, {Watkins}, {Weaver}, {Whitmore},
  {Woillez}, {Zabalza}, \& {Astropy Contributors}}]{astropy:2018}
{Astropy Collaboration}, {Price-Whelan}, A.~M., {Sip{\H{o}}cz}, B.~M., {et~al.}
  2018, \aj, 156, 123, \dodoi{10.3847/1538-3881/aabc4f}

\bibitem[{{Astropy Collaboration} {et~al.}(2022){Astropy Collaboration},
  {Price-Whelan}, {Lim}, {Earl}, {Starkman}, {Bradley}, {Shupe}, {Patil},
  {Corrales}, {Brasseur}, {N{"o}the}, {Donath}, {Tollerud}, {Morris},
  {Ginsburg}, {Vaher}, {Weaver}, {Tocknell}, {Jamieson}, {van Kerkwijk},
  {Robitaille}, {Merry}, {Bachetti}, {G{"u}nther}, {Aldcroft},
  {Alvarado-Montes}, {Archibald}, {B{'o}di}, {Bapat}, {Barentsen}, {Baz{'a}n},
  {Biswas}, {Boquien}, {Burke}, {Cara}, {Cara}, {Conroy}, {Conseil}, {Craig},
  {Cross}, {Cruz}, {D'Eugenio}, {Dencheva}, {Devillepoix}, {Dietrich},
  {Eigenbrot}, {Erben}, {Ferreira}, {Foreman-Mackey}, {Fox}, {Freij}, {Garg},
  {Geda}, {Glattly}, {Gondhalekar}, {Gordon}, {Grant}, {Greenfield}, {Groener},
  {Guest}, {Gurovich}, {Handberg}, {Hart}, {Hatfield-Dodds}, {Homeier},
  {Hosseinzadeh}, {Jenness}, {Jones}, {Joseph}, {Kalmbach}, {Karamehmetoglu},
  {Ka{l}uszy{'n}ski}, {Kelley}, {Kern}, {Kerzendorf}, {Koch}, {Kulumani},
  {Lee}, {Ly}, {Ma}, {MacBride}, {Maljaars}, {Muna}, {Murphy}, {Norman},
  {O'Steen}, {Oman}, {Pacifici}, {Pascual}, {Pascual-Granado}, {Patil},
  {Perren}, {Pickering}, {Rastogi}, {Roulston}, {Ryan}, {Rykoff}, {Sabater},
  {Sakurikar}, {Salgado}, {Sanghi}, {Saunders}, {Savchenko}, {Schwardt},
  {Seifert-Eckert}, {Shih}, {Jain}, {Shukla}, {Sick}, {Simpson},
  {Singanamalla}, {Singer}, {Singhal}, {Sinha}, {Sip{H{o}}cz}, {Spitler},
  {Stansby}, {Streicher}, {{{S}}umak}, {Swinbank}, {Taranu}, {Tewary},
  {Tremblay}, {Val-Borro}, {Van Kooten}, {Vasovi{'c}}, {Verma}, {de Miranda
  Cardoso}, {Williams}, {Wilson}, {Winkel}, {Wood-Vasey}, {Xue}, {Yoachim},
  {Zhang}, {Zonca}, \& {Astropy Project Contributors}}]{astropy:2022}
{Astropy Collaboration}, {Price-Whelan}, A.~M., {Lim}, P.~L., {et~al.} 2022,
  \apj, 935, 167, \dodoi{10.3847/1538-4357/ac7c74}

\bibitem[{{Aungwerojwit} {et~al.}(2024){Aungwerojwit}, {G{\"a}nsicke},
  {Dhillon}, {Drake}, {Inight}, {Kaye}, {Marsh}, {Mullen}, {Pelisoli}, \&
  {Swan}}]{Aungwerojwit2024}
{Aungwerojwit}, A., {G{\"a}nsicke}, B.~T., {Dhillon}, V.~S., {et~al.} 2024,
  \mnras, 530, 117, \dodoi{10.1093/mnras/stae750}

\bibitem[{{Bailer-Jones} {et~al.}(2021){Bailer-Jones}, {Rybizki}, {Fouesneau},
  {Demleitner}, \& {Andrae}}]{BailerJones2021}
{Bailer-Jones}, C.~A.~L., {Rybizki}, J., {Fouesneau}, M., {Demleitner}, M., \&
  {Andrae}, R. 2021, \aj, 161, 147, \dodoi{10.3847/1538-3881/abd806}

\bibitem[{{Ballering} {et~al.}(2022){Ballering}, {Levens}, {Su}, \&
  {Cleeves}}]{Ballering2022}
{Ballering}, N.~P., {Levens}, C.~I., {Su}, K. Y.~L., \& {Cleeves}, L.~I. 2022,
  \apj, 939, 108, \dodoi{10.3847/1538-4357/ac9a4a}

\bibitem[{{Baluev}(2008)}]{Baluev2008}
{Baluev}, R.~V. 2008, \mnras, 385, 1279,
  \dodoi{10.1111/j.1365-2966.2008.12689.x}

\bibitem[{Barbary(2016)}]{exctinction}
Barbary, K. 2016, extinction v0.3.0,  Zenodo, \dodoi{10.5281/zenodo.804967}

\bibitem[{{B{\'e}dard} {et~al.}(2020){B{\'e}dard}, {Bergeron}, {Brassard}, \&
  {Fontaine}}]{Bedard2020}
{B{\'e}dard}, A., {Bergeron}, P., {Brassard}, P., \& {Fontaine}, G. 2020, \apj,
  901, 93, \dodoi{10.3847/1538-4357/abafbe}

\bibitem[{{Beers}(1990)}]{Beers1990}
{Beers}, T.~C. 1990, \aj, 99, 323, \dodoi{10.1086/115330}

\bibitem[{{Bell}(2022)}]{Pyriod2022}
{Bell}, K. 2022, {Pyriod: Period detection and fitting routines}, Astrophysics
  Source Code Library, record ascl:2207.007

\bibitem[{{Bellm} {et~al.}(2019){Bellm}, {Kulkarni}, {Graham}, {Dekany},
  {Smith}, {Riddle}, {Masci}, {Helou}, {Prince}, {Adams}, {Barbarino},
  {Barlow}, {Bauer}, {Beck}, {Belicki}, {Biswas}, {Blagorodnova}, {Bodewits},
  {Bolin}, {Brinnel}, {Brooke}, {Bue}, {Bulla}, {Burruss}, {Cenko}, {Chang},
  {Connolly}, {Coughlin}, {Cromer}, {Cunningham}, {De}, {Delacroix}, {Desai},
  {Duev}, {Eadie}, {Farnham}, {Feeney}, {Feindt}, {Flynn}, {Franckowiak},
  {Frederick}, {Fremling}, {Gal-Yam}, {Gezari}, {Giomi}, {Goldstein},
  {Golkhou}, {Goobar}, {Groom}, {Hacopians}, {Hale}, {Henning}, {Ho}, {Hover},
  {Howell}, {Hung}, {Huppenkothen}, {Imel}, {Ip}, {Ivezi{\'c}}, {Jackson},
  {Jones}, {Juric}, {Kasliwal}, {Kaspi}, {Kaye}, {Kelley}, {Kowalski},
  {Kramer}, {Kupfer}, {Landry}, {Laher}, {Lee}, {Lin}, {Lin}, {Lunnan},
  {Giomi}, {Mahabal}, {Mao}, {Miller}, {Monkewitz}, {Murphy}, {Ngeow},
  {Nordin}, {Nugent}, {Ofek}, {Patterson}, {Penprase}, {Porter}, {Rauch},
  {Rebbapragada}, {Reiley}, {Rigault}, {Rodriguez}, {van Roestel}, {Rusholme},
  {van Santen}, {Schulze}, {Shupe}, {Singer}, {Soumagnac}, {Stein}, {Surace},
  {Sollerman}, {Szkody}, {Taddia}, {Terek}, {Van Sistine}, {van Velzen},
  {Vestrand}, {Walters}, {Ward}, {Ye}, {Yu}, {Yan}, \& {Zolkower}}]{Bellm2019}
{Bellm}, E.~C., {Kulkarni}, S.~R., {Graham}, M.~J., {et~al.} 2019, \pasp, 131,
  018002, \dodoi{10.1088/1538-3873/aaecbe}

\bibitem[{{Bhattacharjee}(2025)}]{Bhattacharjee2025b}
{Bhattacharjee}, S. 2025, arXiv e-prints, arXiv:2507.20594,
  \dodoi{10.48550/arXiv.2507.20594}

\bibitem[{{Bhattacharjee} {et~al.}(2025){Bhattacharjee}, {Vanderbosch},
  {Hollands}, {Tremblay}, {Xu}, {Guidry}, {Hermes}, {Caiazzo}, {Rodriguez},
  {van Roestel}, {Roulston}, {Riddle}, {Rusholme}, {Groom}, {Smith}, \&
  {Toloza}}]{Bhattacharjee2025}
{Bhattacharjee}, S., {Vanderbosch}, Z.~P., {Hollands}, M.~A., {et~al.} 2025,
  arXiv e-prints, arXiv:2502.05502, \dodoi{10.48550/arXiv.2502.05502}

\bibitem[{{Bonsor} {et~al.}(2020){Bonsor}, {Carter}, {Hollands},
  {G{\"a}nsicke}, {Leinhardt}, \& {Harrison}}]{Bonsor2020}
{Bonsor}, A., {Carter}, P.~J., {Hollands}, M., {et~al.} 2020, \mnras, 492,
  2683, \dodoi{10.1093/mnras/stz3603}

\bibitem[{Bradley {et~al.}(2024)Bradley, Sip{\H o}cz, Robitaille, Tollerud,
  Vin{\'{\i}}cius, Deil, Barbary, Wilson, Busko, Donath, G{\"u}nther, Cara,
  Lim, Me{\ss}linger, Conseil, Burnett, Bostroem, Droettboom, Bray, Bratholm,
  Ginsburg, Jamieson, Barentsen, Craig, Morris, Perrin, Rathi, Pascual, \&
  Georgiev}]{larry_bradley_2024_13989456}
Bradley, L., Sip{\H o}cz, B., Robitaille, T., {et~al.} 2024, astropy/photutils:
  2.0.2, 2.0.2,  Zenodo, \dodoi{10.5281/zenodo.13989456}

\bibitem[{{Breger} {et~al.}(1993){Breger}, {Stich}, {Garrido}, {Martin},
  {Jiang}, {Li}, {Hube}, {Ostermann}, {Paparo}, \& {Scheck}}]{Breger1993}
{Breger}, M., {Stich}, J., {Garrido}, R., {et~al.} 1993, \aap, 271, 482

\bibitem[{{Brouwers} {et~al.}(2022){Brouwers}, {Bonsor}, \&
  {Malamud}}]{Brouwers2022}
{Brouwers}, M.~G., {Bonsor}, A., \& {Malamud}, U. 2022, \mnras, 509, 2404,
  \dodoi{10.1093/mnras/stab3009}

\bibitem[{{Cauley} {et~al.}(2018){Cauley}, {Farihi}, {Redfield}, {Bachman},
  {Parsons}, \& {G{\"a}nsicke}}]{Cauley2018}
{Cauley}, P.~W., {Farihi}, J., {Redfield}, S., {et~al.} 2018, \apjl, 852, L22,
  \dodoi{10.3847/2041-8213/aaa3d9}

\bibitem[{{Coutu} {et~al.}(2019){Coutu}, {Dufour}, {Bergeron}, {Blouin},
  {Loranger}, {Allard}, \& {Dunlap}}]{Coutu2019}
{Coutu}, S., {Dufour}, P., {Bergeron}, P., {et~al.} 2019, \apj, 885, 74,
  \dodoi{10.3847/1538-4357/ab46b9}

\bibitem[{Craig {et~al.}(2017{\natexlab{a}})Craig, Crawford, Seifert,
  Robitaille, Sip{\H o}cz, Walawender, Vin{\'{\i}}cius, Ninan, Droettboom,
  Youn, Tollerud, Bray, Walker, Janga, Stotts, G{\"u}nther, Rol, Bach, Bradley,
  Deil, Price-Whelan, Barbary, Horton, Schoenell, Heidt, Gasdia, Nelson, \&
  Streicher}]{ccdproc_2017}
Craig, M., Crawford, S., Seifert, M., {et~al.} 2017{\natexlab{a}},
  astropy/ccdproc: v1.3.0.post1, \dodoi{10.5281/zenodo.1069648}

\bibitem[{Craig {et~al.}(2017{\natexlab{b}})Craig, Crawford, Seifert,
  Robitaille, Sip{\H o}cz, Walawender, Vin{\'{\i}}cius, Ninan, Droettboom,
  Youn, Tollerud, Bray, Walker, Janga, Stotts, G{\"u}nther, Rol, Bach, Bradley,
  Deil, Price-Whelan, Barbary, Horton, Schoenell, Heidt, Gasdia, Nelson, \&
  Streicher}]{matt_craig_2017_1069648}
---. 2017{\natexlab{b}}, astropy/ccdproc: v1.3.0.post1,
  \dodoi{10.5281/zenodo.1069648}

\bibitem[{{Croll} {et~al.}(2017){Croll}, {Dalba}, {Vanderburg}, {Eastman},
  {Rappaport}, {DeVore}, {Bieryla}, {Muirhead}, {Han}, {Latham}, {Beatty},
  {Wittenmyer}, {Wright}, {Johnson}, \& {McCrady}}]{Croll2017}
{Croll}, B., {Dalba}, P.~A., {Vanderburg}, A., {et~al.} 2017, \apj, 836, 82,
  \dodoi{10.3847/1538-4357/836/1/82}

\bibitem[{{Cunningham} {et~al.}(2021){Cunningham}, {Tremblay}, {Bauer},
  {Toloza}, {Cukanovaite}, {Koester}, {Farihi}, {Freytag}, {G{\"a}nsicke},
  {Ludwig}, \& {Veras}}]{Cunningham2021}
{Cunningham}, T., {Tremblay}, P.-E., {Bauer}, E.~B., {et~al.} 2021, \mnras,
  503, 1646, \dodoi{10.1093/mnras/stab553}

\bibitem[{{Debes} {et~al.}(2012){Debes}, {Kilic}, {Faedi}, {Shkolnik},
  {Lopez-Morales}, {Weinberger}, {Slesnick}, \& {West}}]{Debes2012}
{Debes}, J.~H., {Kilic}, M., {Faedi}, F., {et~al.} 2012, \apj, 754, 59,
  \dodoi{10.1088/0004-637X/754/1/59}

\bibitem[{{Debes} \& {Sigurdsson}(2002)}]{Debes2002}
{Debes}, J.~H., \& {Sigurdsson}, S. 2002, \apj, 572, 556,
  \dodoi{10.1086/340291}

\bibitem[{{Dennihy} {et~al.}(2020){Dennihy}, {Farihi}, {Gentile Fusillo}, \&
  {Debes}}]{Dennihy2020a}
{Dennihy}, E., {Farihi}, J., {Gentile Fusillo}, N.~P., \& {Debes}, J.~H. 2020,
  \apj, 891, 97, \dodoi{10.3847/1538-4357/ab7249}

\bibitem[{{Dhillon} {et~al.}(2021){Dhillon}, {Bezawada}, {Black}, {Dixon},
  {Gamble}, {Gao}, {Henry}, {Kerry}, {Littlefair}, {Lunney}, {Marsh}, {Miller},
  {Parsons}, {Ashley}, {Breedt}, {Brown}, {Dyer}, {Green}, {Pelisoli},
  {Sahman}, {Wild}, {Ives}, {Mehrgan}, {Stegmeier}, {Dubbeldam}, {Morris},
  {Osborn}, {Wilson}, {Casares}, {Mu{\~n}oz-Darias}, {Pall{\'e}},
  {Rodr{\'\i}guez-Gil}, {Shahbaz}, {Torres}, {de Ugarte Postigo},
  {Cabrera-Lavers}, {Corradi}, {Dom{\'\i}nguez}, \&
  {Garc{\'\i}a-Alvarez}}]{Dhillon2021}
{Dhillon}, V.~S., {Bezawada}, N., {Black}, M., {et~al.} 2021, \mnras, 507, 350,
  \dodoi{10.1093/mnras/stab2130}

\bibitem[{{Draine} \& {Lee}(1984)}]{Draine&Lee1984}
{Draine}, B.~T., \& {Lee}, H.~M. 1984, \apj, 285, 89, \dodoi{10.1086/162480}

\bibitem[{{El-Badry} {et~al.}(2022){El-Badry}, {Conroy}, {Fuller}, {Kiman},
  {van Roestel}, {Rodriguez}, \& {Burdge}}]{El-Badry2022}
{El-Badry}, K., {Conroy}, C., {Fuller}, J., {et~al.} 2022, \mnras, 517, 4916,
  \dodoi{10.1093/mnras/stac2945}

\bibitem[{{Farihi}(2016)}]{Farihi2016}
{Farihi}, J. 2016, \nar, 71, 9, \dodoi{10.1016/j.newar.2016.03.001}

\bibitem[{{Farihi} {et~al.}(2022){Farihi}, {Hermes}, {Marsh}, {Mustill},
  {Wyatt}, {Guidry}, {Wilson}, {Redfield}, {Izquierdo}, {Toloza},
  {G{\"a}nsicke}, {Aungwerojwit}, {Kaewmanee}, {Dhillon}, \&
  {Swan}}]{Farihi2022}
{Farihi}, J., {Hermes}, J.~J., {Marsh}, T.~R., {et~al.} 2022, \mnras, 511,
  1647, \dodoi{10.1093/mnras/stab3475}

\bibitem[{{Fortin-Archambault} {et~al.}(2020){Fortin-Archambault}, {Dufour}, \&
  {Xu}}]{Fortin-Archambault2020}
{Fortin-Archambault}, M., {Dufour}, P., \& {Xu}, S. 2020, \apj, 888, 47,
  \dodoi{10.3847/1538-4357/ab585a}

\bibitem[{{Fr{\"o}hlich} \& {Reg{\'a}ly}(2024)}]{2024A&A...692A..25F}
{Fr{\"o}hlich}, V., \& {Reg{\'a}ly}, Z. 2024, \aap, 692, A25,
  \dodoi{10.1051/0004-6361/202450471}

\bibitem[{{G{\"a}nsicke} {et~al.}(2012){G{\"a}nsicke}, {Koester}, {Farihi},
  {Girven}, {Parsons}, \& {Breedt}}]{Gaensicke2012}
{G{\"a}nsicke}, B.~T., {Koester}, D., {Farihi}, J., {et~al.} 2012, \mnras, 424,
  333, \dodoi{10.1111/j.1365-2966.2012.21201.x}

\bibitem[{{G{\"a}nsicke} {et~al.}(2016){G{\"a}nsicke}, {Aungwerojwit}, {Marsh},
  {Dhillon}, {Sahman}, {Veras}, {Farihi}, {Chote}, {Ashley}, {Arjyotha},
  {Rattanasoon}, {Littlefair}, {Pollacco}, \& {Burleigh}}]{Gaensicke2016}
{G{\"a}nsicke}, B.~T., {Aungwerojwit}, A., {Marsh}, T.~R., {et~al.} 2016,
  \apjl, 818, L7, \dodoi{10.3847/2041-8205/818/1/L7}

\bibitem[{{Gary} {et~al.}(2017){Gary}, {Rappaport}, {Kaye}, {Alonso}, \&
  {Hambschs}}]{Gary2017}
{Gary}, B.~L., {Rappaport}, S., {Kaye}, T.~G., {Alonso}, R., \& {Hambschs},
  F.~J. 2017, \mnras, 465, 3267, \dodoi{10.1093/mnras/stw2921}

\bibitem[{{Gentile Fusillo} {et~al.}(2021){Gentile Fusillo}, {Tremblay},
  {Cukanovaite}, {Vorontseva}, {Lallement}, {Hollands}, {G{\"a}nsicke},
  {Burdge}, {McCleery}, \& {Jordan}}]{GentileFusillo2021}
{Gentile Fusillo}, N.~P., {Tremblay}, P.~E., {Cukanovaite}, E., {et~al.} 2021,
  \mnras, 508, 3877, \dodoi{10.1093/mnras/stab2672}

\bibitem[{{Ginsburg} {et~al.}(2019){Ginsburg}, {Sip{\H{o}}cz}, {Brasseur},
  {Cowperthwaite}, {Craig}, {Deil}, {Guillochon}, {Guzman}, {Liedtke}, {Lian
  Lim}, {Lockhart}, {Mommert}, {Morris}, {Norman}, {Parikh}, {Persson},
  {Robitaille}, {Segovia}, {Singer}, {Tollerud}, {de Val-Borro}, {Valtchanov},
  {Woillez}, {Astroquery Collaboration}, \& {a subset of astropy
  Collaboration}}]{2019AJ....157...98G}
{Ginsburg}, A., {Sip{\H{o}}cz}, B.~M., {Brasseur}, C.~E., {et~al.} 2019, \aj,
  157, 98, \dodoi{10.3847/1538-3881/aafc33}

\bibitem[{{Girven} {et~al.}(2012){Girven}, {Brinkworth}, {Farihi},
  {G{\"a}nsicke}, {Hoard}, {Marsh}, \& {Koester}}]{Girven2012}
{Girven}, J., {Brinkworth}, C.~S., {Farihi}, J., {et~al.} 2012, \apj, 749, 154,
  \dodoi{10.1088/0004-637X/749/2/154}

\bibitem[{{Graham} {et~al.}(2013){Graham}, {Drake}, {Djorgovski}, {Mahabal}, \&
  {Donalek}}]{Graham2013_CE}
{Graham}, M.~J., {Drake}, A.~J., {Djorgovski}, S.~G., {Mahabal}, A.~A., \&
  {Donalek}, C. 2013, \mnras, 434, 2629, \dodoi{10.1093/mnras/stt1206}

\bibitem[{{Green} {et~al.}(2015){Green}, {Schlafly}, {Finkbeiner}, {Rix},
  {Martin}, {Burgett}, {Draper}, {Flewelling}, {Hodapp}, {Kaiser}, {Kudritzki},
  {Magnier}, {Metcalfe}, {Price}, {Tonry}, \& {Wainscoat}}]{Green2015}
{Green}, G.~M., {Schlafly}, E.~F., {Finkbeiner}, D.~P., {et~al.} 2015, \apj,
  810, 25, \dodoi{10.1088/0004-637X/810/1/25}

\bibitem[{{Green} {et~al.}(2018){Green}, {Schlafly}, {Finkbeiner}, {Rix},
  {Martin}, {Burgett}, {Draper}, {Flewelling}, {Hodapp}, {Kaiser}, {Kudritzki},
  {Magnier}, {Metcalfe}, {Tonry}, {Wainscoat}, \& {Waters}}]{Green2018}
{Green}, G.~M., {Schlafly}, E.~F., {Finkbeiner}, D., {et~al.} 2018, \mnras,
  478, 651, \dodoi{10.1093/mnras/sty1008}

\bibitem[{{Guidry} {et~al.}(2024){Guidry}, {Hermes}, {De}, {Ould Rouis},
  {Ewing}, \& {Kaiser}}]{Guidry2024}
{Guidry}, J.~A., {Hermes}, J.~J., {De}, K., {et~al.} 2024, \apj, 972, 126,
  \dodoi{10.3847/1538-4357/ad5be7}

\bibitem[{{Guidry} {et~al.}(2021){Guidry}, {Vanderbosch}, {Hermes}, {Barlow},
  {Lopez}, {Boudreaux}, {Corcoran}, {Bell}, {Montgomery}, {Heintz},
  {Castanheira}, {Reding}, {Dunlap}, {Winget}, {Winget}, \&
  {Kuehne}}]{Guidry2021}
{Guidry}, J.~A., {Vanderbosch}, Z.~P., {Hermes}, J.~J., {et~al.} 2021, \apj,
  912, 125, \dodoi{10.3847/1538-4357/abee68}

\bibitem[{{Hallakoun} {et~al.}(2017){Hallakoun}, {Xu}, {Maoz}, {Marsh},
  {Ivanov}, {Dhillon}, {Bours}, {Parsons}, {Kerry}, {Sharma}, {Su},
  {Rengaswamy}, {Pravec}, {Ku{\v{s}}nir{\'a}k}, {Ku{\v{c}}{\'a}kov{\'a}},
  {Armstrong}, {Arnold}, {Gerard}, \& {Vanzi}}]{Hallakoun2017}
{Hallakoun}, N., {Xu}, S., {Maoz}, D., {et~al.} 2017, \mnras, 469, 3213,
  \dodoi{10.1093/mnras/stx924}

\bibitem[{{Hara} \& {Ford}(2023)}]{Hara&Ford2023}
{Hara}, N.~C., \& {Ford}, E.~B. 2023, Annual Review of Statistics and Its
  Application, 10, 623, \dodoi{10.1146/annurev-statistics-033021-012225}

\bibitem[{{Harding} {et~al.}(2016){Harding}, {Hallinan}, {Milburn}, {Gardner},
  {Konidaris}, {Singh}, {Shao}, {Sandhu}, {Kyne}, \&
  {Schlichting}}]{Harding2016}
{Harding}, L.~K., {Hallinan}, G., {Milburn}, J., {et~al.} 2016, \mnras, 457,
  3036, \dodoi{10.1093/mnras/stw094}

\bibitem[{Harris {et~al.}(2020)Harris, Millman, van~der Walt, Gommers,
  Virtanen, Cournapeau, Wieser, Taylor, Berg, Smith, Kern, Picus, Hoyer, van
  Kerkwijk, Brett, Haldane, del R{\'{i}}o, Wiebe, Peterson,
  G{\'{e}}rard-Marchant, Sheppard, Reddy, Weckesser, Abbasi, Gohlke, \&
  Oliphant}]{harris2020array}
Harris, C.~R., Millman, K.~J., van~der Walt, S.~J., {et~al.} 2020, Nature, 585,
  357, \dodoi{10.1038/s41586-020-2649-2}

\bibitem[{{Hermes} {et~al.}(2025){Hermes}, {Guidry}, {Vanderbosch},
  {Badenas-Agusti}, {Xu}, {Kao}, {Rodriguez}, \& {Hawkins}}]{Hermes2025}
{Hermes}, J.~J., {Guidry}, J.~A., {Vanderbosch}, Z.~P., {et~al.} 2025, \apj,
  980, 56, \dodoi{10.3847/1538-4357/ada5fd}

\bibitem[{{Hoffman}(2022)}]{cuvarbase2022}
{Hoffman}, J. 2022, {cuvarbase: fast period finding utilities for GPUs},
  Astrophysics Source Code Library, record ascl:2210.030

\bibitem[{Hunter(2007)}]{Hunter:2007}
Hunter, J.~D. 2007, Computing in Science \& Engineering, 9, 90,
  \dodoi{10.1109/MCSE.2007.55}

\bibitem[{{Ivezi{\'c}} {et~al.}(2019){Ivezi{\'c}}, {Kahn}, {Tyson}, {Abel},
  {Acosta}, {Allsman}, {Alonso}, {AlSayyad}, {Anderson}, {Andrew}, \&
  et~al.}]{2019ApJ...873..111I}
{Ivezi{\'c}}, {\v{Z}}., {Kahn}, S.~M., {Tyson}, J.~A., {et~al.} 2019, \apj,
  873, 111, \dodoi{10.3847/1538-4357/ab042c}

\bibitem[{{Izquierdo} {et~al.}(2018){Izquierdo}, {Rodr{\'\i}guez-Gil},
  {G{\"a}nsicke}, {Mustill}, {Toloza}, {Tremblay}, {Wyatt}, {Chote}, {Eggl},
  {Farihi}, {Koester}, {Lyra}, {Manser}, {Marsh}, {Pall{\'e}}, {Raddi},
  {Veras}, {Villaver}, \& {Portegies Zwart}}]{Izquierdo2018}
{Izquierdo}, P., {Rodr{\'\i}guez-Gil}, P., {G{\"a}nsicke}, B.~T., {et~al.}
  2018, \mnras, 481, 703, \dodoi{10.1093/mnras/sty2315}

\bibitem[{{Janes} {et~al.}(2004){Janes}, {Clemens}, {Hayes-Gehrke}, {Eastman},
  {Sarcia}, \& {Bosh}}]{Janes2004}
{Janes}, K.~A., {Clemens}, D.~P., {Hayes-Gehrke}, M.~N., {et~al.} 2004, in
  American Astronomical Society Meeting Abstracts, Vol. 204, American
  Astronomical Society Meeting Abstracts \#204, 10.01

\bibitem[{{Jura}(2003)}]{Jura2003}
{Jura}, M. 2003, \apjl, 584, L91, \dodoi{10.1086/374036}

\bibitem[{{Karjalainen} {et~al.}(2019){Karjalainen}, {de Mooij}, {Karjalainen},
  \& {Gibson}}]{Karjalainen2019}
{Karjalainen}, M., {de Mooij}, E. J.~W., {Karjalainen}, R., \& {Gibson}, N.~P.
  2019, \mnras, 482, 999, \dodoi{10.1093/mnras/sty2778}

\bibitem[{{Katz} {et~al.}(2021){Katz}, {Cooper}, {Coughlin}, {Burdge},
  {Breivik}, \& {Larson}}]{Katz2021}
{Katz}, M.~L., {Cooper}, O.~R., {Coughlin}, M.~W., {et~al.} 2021, \mnras, 503,
  2665, \dodoi{10.1093/mnras/stab504}

\bibitem[{{Kenyon} \& {Bromley}(2017{\natexlab{a}})}]{KenyonBromley2017a}
{Kenyon}, S.~J., \& {Bromley}, B.~C. 2017{\natexlab{a}}, \apj, 844, 116,
  \dodoi{10.3847/1538-4357/aa7b85}

\bibitem[{{Kenyon} \& {Bromley}(2017{\natexlab{b}})}]{KenyonBromley2017b}
---. 2017{\natexlab{b}}, \apj, 850, 50, \dodoi{10.3847/1538-4357/aa9570}

\bibitem[{{Kim} \& {Martin}(1995)}]{Kim1995}
{Kim}, S.-H., \& {Martin}, P.~G. 1995, \apj, 444, 293, \dodoi{10.1086/175604}

\bibitem[{{Kobayashi} {et~al.}(2011){Kobayashi}, {Kimura}, {Watanabe},
  {Yamamoto}, \& {M{\"u}ller}}]{2011EP&S...63.1067K}
{Kobayashi}, H., {Kimura}, H., {Watanabe}, S.-i., {Yamamoto}, T., \&
  {M{\"u}ller}, S. 2011, Earth, Planets and Space, 63, 1067,
  \dodoi{10.5047/eps.2011.03.012}

\bibitem[{{Koester}(2010)}]{Koester2010}
{Koester}, D. 2010, \memsai, 81, 921

\bibitem[{{Koester} {et~al.}(2014){Koester}, {G{\"a}nsicke}, \&
  {Farihi}}]{Koester2014}
{Koester}, D., {G{\"a}nsicke}, B.~T., \& {Farihi}, J. 2014, \aap, 566, A34,
  \dodoi{10.1051/0004-6361/201423691}

\bibitem[{{Kuschnig} {et~al.}(1997){Kuschnig}, {Weiss}, {Gruber}, {Bely}, \&
  {Jenkner}}]{Kuschnig1997}
{Kuschnig}, R., {Weiss}, W.~W., {Gruber}, R., {Bely}, P.~Y., \& {Jenkner}, H.
  1997, \aap, 328, 544

\bibitem[{{Le Bourdais} {et~al.}(2024){Le Bourdais}, {Dufour}, \&
  {Xu}}]{LeBourdais2024}
{Le Bourdais}, {\'E}., {Dufour}, P., \& {Xu}, S. 2024, \apj, 977, 93,
  \dodoi{10.3847/1538-4357/ad90b7}

\bibitem[{{Li} {et~al.}(2025{\natexlab{a}}){Li}, {Bonsor}, \&
  {Shorttle}}]{Li2025b}
{Li}, Y., {Bonsor}, A., \& {Shorttle}, O. 2025{\natexlab{a}}, \mnras, 541, 610,
  \dodoi{10.1093/mnras/staf1028}

\bibitem[{{Li} {et~al.}(2025{\natexlab{b}}){Li}, {Bonsor}, {Shorttle}, \&
  {Rogers}}]{Li2025a}
{Li}, Y., {Bonsor}, A., {Shorttle}, O., \& {Rogers}, L.~K. 2025{\natexlab{b}},
  \mnras, 537, 2214, \dodoi{10.1093/mnras/staf182}

\bibitem[{{Lomb}(1976)}]{Lomb1976}
{Lomb}, N.~R. 1976, \apss, 39, 447, \dodoi{10.1007/BF00648343}

\bibitem[{{Manser} {et~al.}(2020){Manser}, {G{\"a}nsicke}, {Gentile Fusillo},
  {Ashley}, {Breedt}, {Hollands}, {Izquierdo}, \& {Pelisoli}}]{Manser2020}
{Manser}, C.~J., {G{\"a}nsicke}, B.~T., {Gentile Fusillo}, N.~P., {et~al.}
  2020, \mnras, 493, 2127, \dodoi{10.1093/mnras/staa359}

\bibitem[{{Marocco} {et~al.}(2021){Marocco}, {Eisenhardt}, {Fowler},
  {Kirkpatrick}, {Meisner}, {Schlafly}, {Stanford}, {Garcia}, {Caselden},
  {Cushing}, {Cutri}, {Faherty}, {Gelino}, {Gonzalez}, {Jarrett}, {Koontz},
  {Mainzer}, {Marchese}, {Mobasher}, {Schlegel}, {Stern}, {Teplitz}, \&
  {Wright}}]{Marocco2021}
{Marocco}, F., {Eisenhardt}, P. R.~M., {Fowler}, J.~W., {et~al.} 2021, \apjs,
  253, 8, \dodoi{10.3847/1538-4365/abd805}

\bibitem[{{Marrese} {et~al.}(2022){Marrese}, {Marinoni}, {Fabrizio}, \&
  {Altavilla}}]{Marrese2022}
{Marrese}, P.~M., {Marinoni}, S., {Fabrizio}, M., \& {Altavilla}, G. 2022,
  {Gaia DR3 documentation Chapter 15: Cross-match with external catalogues},
  Gaia DR3 documentation, European Space Agency; Gaia Data Processing and
  Analysis Consortium.

\bibitem[{{Masci} {et~al.}(2019){Masci}, {Laher}, {Rusholme}, {Shupe}, {Groom},
  {Surace}, {Jackson}, {Monkewitz}, {Beck}, \& {Flynn}}]{Masci2019}
{Masci}, F.~J., {Laher}, R.~R., {Rusholme}, B., {et~al.} 2019, \pasp, 131,
  018003, \dodoi{10.1088/1538-3873/aae8ac}

\bibitem[{{Masci} {et~al.}(2023){Masci}, {Laher}, {Rusholme}, {Shupe},
  {Paladini}, {Groom}, {Wold}, {Miller}, \& {Drake}}]{Masci2023}
---. 2023, arXiv e-prints, arXiv:2305.16279, \dodoi{10.48550/arXiv.2305.16279}

\bibitem[{{Mathis} {et~al.}(1977){Mathis}, {Rumpl}, \&
  {Nordsieck}}]{Mathis1977}
{Mathis}, J.~S., {Rumpl}, W., \& {Nordsieck}, K.~H. 1977, \apj, 217, 425,
  \dodoi{10.1086/155591}

\bibitem[{{McDonald} \& {Veras}(2021)}]{McDonald2021}
{McDonald}, C.~H., \& {Veras}, D. 2021, \mnras, 506, 4031,
  \dodoi{10.1093/mnras/stab1906}

\bibitem[{{Miranda} \& {Rafikov}(2018)}]{Miranda2018}
{Miranda}, R., \& {Rafikov}, R.~R. 2018, \apj, 857, 135,
  \dodoi{10.3847/1538-4357/aab9a2}

\bibitem[{{Montgomery} \& {O'Donoghue}(1999)}]{Montgomery1999}
{Montgomery}, M.~H., \& {O'Donoghue}, D. 1999, Delta Scuti Star Newsletter, 13,
  28

\bibitem[{{Murga} {et~al.}(2015){Murga}, {Zhu}, {M{\'e}nard}, \&
  {Lan}}]{Murga2015}
{Murga}, M., {Zhu}, G., {M{\'e}nard}, B., \& {Lan}, T.-W. 2015, \mnras, 452,
  511, \dodoi{10.1093/mnras/stv1277}

\bibitem[{{Nagarajan} {et~al.}(2023){Nagarajan}, {El-Badry}, {Rodriguez}, {van
  Roestel}, \& {Roulston}}]{Nagarajan2023}
{Nagarajan}, P., {El-Badry}, K., {Rodriguez}, A.~C., {van Roestel}, J., \&
  {Roulston}, B. 2023, \mnras, 524, 4367, \dodoi{10.1093/mnras/stad2130}

\bibitem[{{Nauenberg}(1972)}]{Nauenberg1972}
{Nauenberg}, M. 1972, \apj, 175, 417, \dodoi{10.1086/151568}

\bibitem[{{Nesvold} {et~al.}(2016){Nesvold}, {Naoz}, {Vican}, \&
  {Farr}}]{Nesvold2016}
{Nesvold}, E.~R., {Naoz}, S., {Vican}, L., \& {Farr}, W.~M. 2016, \apj, 826,
  19, \dodoi{10.3847/0004-637X/826/1/19}

\bibitem[{Newville {et~al.}(2014)Newville, Stensitzki, Allen, \&
  Ingargiola}]{LMFIT_2014}
Newville, M., Stensitzki, T., Allen, D.~B., \& Ingargiola, A. 2014, {LMFIT:
  Non-Linear Least-Square Minimization and Curve-Fitting for Python}, 0.8.0,
  Zenodo, \dodoi{10.5281/zenodo.11813}

\bibitem[{{Noor} {et~al.}(2025){Noor}, {Farihi}, {Kenyon}, {Rafikov}, {Wyatt},
  {Su}, {Melis}, {Swan}, {Wilson}, {G{\"a}nsicke}, {Bonsor}, {Rogers},
  {Redfield}, \& {Kilic}}]{Noor2025}
{Noor}, H.~T., {Farihi}, J., {Kenyon}, S.~J., {et~al.} 2025, arXiv e-prints,
  arXiv:2508.13119.
\newblock \doarXiv{2508.13119}

\bibitem[{{O'Brien} {et~al.}(2024){O'Brien}, {Tremblay}, {Klein}, {Koester},
  {Melis}, {B{\'e}dard}, {Cukanovaite}, {Cunningham}, {Doyle}, {G{\"a}nsicke},
  {Gentile Fusillo}, {Hollands}, {McCleery}, {Pelisoli}, {Toonen},
  {Weinberger}, \& {Zuckerman}}]{OBrien2024}
{O'Brien}, M.~W., {Tremblay}, P.~E., {Klein}, B.~L., {et~al.} 2024, \mnras,
  527, 8687, \dodoi{10.1093/mnras/stad3773}

\bibitem[{{O'Connor} {et~al.}(2022){O'Connor}, {Teyssandier}, \&
  {Lai}}]{OConnor2022}
{O'Connor}, C.~E., {Teyssandier}, J., \& {Lai}, D. 2022, \mnras, 513, 4178,
  \dodoi{10.1093/mnras/stac1189}

\bibitem[{{Oke} \& {Gunn}(1982)}]{Oke82}
{Oke}, J.~B., \& {Gunn}, J.~E. 1982, \pasp, 94, 586, \dodoi{10.1086/131027}

\bibitem[{{Oke} {et~al.}(1995){Oke}, {Cohen}, {Carr}, {Cromer}, {Dingizian},
  {Harris}, {Labrecque}, {Lucinio}, {Schaal}, {Epps}, \& {Miller}}]{Oke95}
{Oke}, J.~B., {Cohen}, J.~G., {Carr}, M., {et~al.} 1995, \pasp, 107, 375,
  \dodoi{10.1086/133562}

\bibitem[{{Ould Rouis} {et~al.}(2024){Ould Rouis}, {Hermes}, {G{\"a}nsicke},
  {Sahu}, {Koester}, {Tremblay}, {Veras}, {Farihi}, {Heintz}, {Gentile
  Fusillo}, \& {Redfield}}]{OuldRouis2024}
{Ould Rouis}, L.~B., {Hermes}, J.~J., {G{\"a}nsicke}, B.~T., {et~al.} 2024,
  \apj, 976, 156, \dodoi{10.3847/1538-4357/ad86bb}

\bibitem[{pandas~development team(2024)}]{the_pandas_dev_team_2024_13819579}
pandas~development team, T. 2024, pandas-dev/pandas: Pandas, v2.2.3,  Zenodo,
  \dodoi{10.5281/zenodo.13819579}

\bibitem[{{Perley}(2019)}]{Perley19}
{Perley}, D.~A. 2019, \pasp, 131, 084503, \dodoi{10.1088/1538-3873/ab215d}

\bibitem[{{Poznanski} {et~al.}(2012){Poznanski}, {Prochaska}, \&
  {Bloom}}]{Poznanski2012}
{Poznanski}, D., {Prochaska}, J.~X., \& {Bloom}, J.~S. 2012, \mnras, 426, 1465,
  \dodoi{10.1111/j.1365-2966.2012.21796.x}

\bibitem[{{Rafikov} \& {Garmilla}(2012)}]{Rafikov&Garmilla2012}
{Rafikov}, R.~R., \& {Garmilla}, J.~A. 2012, \apj, 760, 123,
  \dodoi{10.1088/0004-637X/760/2/123}

\bibitem[{{Rappaport} {et~al.}(2016){Rappaport}, {Gary}, {Kaye}, {Vanderburg},
  {Croll}, {Benni}, \& {Foote}}]{Rappaport2016}
{Rappaport}, S., {Gary}, B.~L., {Kaye}, T., {et~al.} 2016, \mnras, 458, 3904,
  \dodoi{10.1093/mnras/stw612}

\bibitem[{{Reach} {et~al.}(2005){Reach}, {Megeath}, {Cohen}, {Hora}, {Carey},
  {Surace}, {Willner}, {Barmby}, {Wilson}, {Glaccum}, {Lowrance}, {Marengo}, \&
  {Fazio}}]{Reach2005}
{Reach}, W.~T., {Megeath}, S.~T., {Cohen}, M., {et~al.} 2005, \pasp, 117, 978,
  \dodoi{10.1086/432670}

\bibitem[{{Redfield} {et~al.}(2017){Redfield}, {Farihi}, {Cauley}, {Parsons},
  {G{\"a}nsicke}, \& {Duvvuri}}]{Redfield2017}
{Redfield}, S., {Farihi}, J., {Cauley}, P.~W., {et~al.} 2017, \apj, 839, 42,
  \dodoi{10.3847/1538-4357/aa68a0}

\bibitem[{{Rocchetto} {et~al.}(2015){Rocchetto}, {Farihi}, {G{\"a}nsicke}, \&
  {Bergfors}}]{Rocchetto2015}
{Rocchetto}, M., {Farihi}, J., {G{\"a}nsicke}, B.~T., \& {Bergfors}, C. 2015,
  \mnras, 449, 574, \dodoi{10.1093/mnras/stv282}

\bibitem[{{Rockosi} {et~al.}(2010){Rockosi}, {Stover}, {Kibrick}, {Lockwood},
  {Peck}, {Cowley}, {Bolte}, {Adkins}, {Alcott}, {Allen}, {Brown}, {Cabak},
  {Deich}, {Hilyard}, {Kassis}, {Lanclos}, {Lewis}, {Pfister}, {Phillips},
  {Robinson}, {Saylor}, {Thompson}, {Ward}, {Wei}, \& {Wright}}]{Rockosi10}
{Rockosi}, C., {Stover}, R., {Kibrick}, R., {et~al.} 2010, in Society of
  Photo-Optical Instrumentation Engineers (SPIE) Conference Series, Vol. 7735,
  Ground-based and Airborne Instrumentation for Astronomy III, ed. I.~S.
  {McLean}, S.~K. {Ramsay}, \& H.~{Takami}, 77350R, \dodoi{10.1117/12.856818}

\bibitem[{{Rogers} {et~al.}(2024){Rogers}, {Bonsor}, {Xu}, {Buchan}, {Dufour},
  {Klein}, {Hodgkin}, {Kissler-Patig}, {Melis}, {Walton}, \&
  {Weinberger}}]{Rogers2024}
{Rogers}, L.~K., {Bonsor}, A., {Xu}, S., {et~al.} 2024, \mnras, 532, 3866,
  \dodoi{10.1093/mnras/stae1520}

\bibitem[{{Scargle}(1982)}]{Scargle1982}
{Scargle}, J.~D. 1982, \apj, 263, \dodoi{10.1086/160554}

\bibitem[{{Schlafly} {et~al.}(2019){Schlafly}, {Meisner}, \&
  {Green}}]{Schlafly2019}
{Schlafly}, E.~F., {Meisner}, A.~M., \& {Green}, G.~M. 2019, \apjs, 240, 30,
  \dodoi{10.3847/1538-4365/aafbea}

\bibitem[{{Swan} {et~al.}(2020){Swan}, {Farihi}, {Wilson}, \&
  {Parsons}}]{Swan2020}
{Swan}, A., {Farihi}, J., {Wilson}, T.~G., \& {Parsons}, S.~G. 2020, \mnras,
  496, 5233, \dodoi{10.1093/mnras/staa1688}

\bibitem[{{Swan} {et~al.}(2021){Swan}, {Kenyon}, {Farihi}, {Dennihy},
  {G{\"a}nsicke}, {Hermes}, {Melis}, \& {von Hippel}}]{Swan2021}
{Swan}, A., {Kenyon}, S.~J., {Farihi}, J., {et~al.} 2021, \mnras, 506, 432,
  \dodoi{10.1093/mnras/stab1738}

\bibitem[{{Tamburo} {et~al.}(2023){Tamburo}, {Muirhead}, \&
  {Dressing}}]{Tamburo2023}
{Tamburo}, P., {Muirhead}, P.~S., \& {Dressing}, C.~D. 2023, \aj, 165, 251,
  \dodoi{10.3847/1538-3881/acd1de}

\bibitem[{{van Lieshout} {et~al.}(2018){van Lieshout}, {Kral}, {Charnoz},
  {Wyatt}, \& {Shannon}}]{vanLieshout2018}
{van Lieshout}, R., {Kral}, Q., {Charnoz}, S., {Wyatt}, M.~C., \& {Shannon}, A.
  2018, \mnras, 480, 2784, \dodoi{10.1093/mnras/sty1271}

\bibitem[{{van Lieshout} {et~al.}(2014){van Lieshout}, {Min}, \&
  {Dominik}}]{vanLieshout2014}
{van Lieshout}, R., {Min}, M., \& {Dominik}, C. 2014, \aap, 572, A76,
  \dodoi{10.1051/0004-6361/201424876}

\bibitem[{Vanderbosch(2023)}]{phot2lc}
Vanderbosch, Z. 2023, zvanderbosch/phot2lc: phot2lc v1.7.8 release, v1.7.8,
  Zenodo, \dodoi{10.5281/zenodo.8169807}

\bibitem[{{Vanderbosch} {et~al.}(2020){Vanderbosch}, {Hermes}, {Dennihy},
  {Dunlap}, {Izquierdo}, {Tremblay}, {Cho}, {G{\"a}nsicke}, {Toloza}, {Bell},
  {Montgomery}, \& {Winget}}]{Vanderbosch2020}
{Vanderbosch}, Z., {Hermes}, J.~J., {Dennihy}, E., {et~al.} 2020, \apj, 897,
  171, \dodoi{10.3847/1538-4357/ab9649}

\bibitem[{{Vanderbosch} {et~al.}(2021){Vanderbosch}, {Rappaport}, {Guidry},
  {Gary}, {Blouin}, {Kaye}, {Weinberger}, {Melis}, {Klein}, {Zuckerman},
  {Vanderburg}, {Hermes}, {Hegedus}, {Burleigh}, {Sefako}, {Worters}, \&
  {Heintz}}]{Vanderbosch2021}
{Vanderbosch}, Z.~P., {Rappaport}, S., {Guidry}, J.~A., {et~al.} 2021, \apj,
  917, 41, \dodoi{10.3847/1538-4357/ac0822}

\bibitem[{{Vanderburg} {et~al.}(2015){Vanderburg}, {Johnson}, {Rappaport},
  {Bieryla}, {Irwin}, {Lewis}, {Kipping}, {Brown}, {Dufour}, {Ciardi}, {Angus},
  {Schaefer}, {Latham}, {Charbonneau}, {Beichman}, {Eastman}, {McCrady},
  {Wittenmyer}, \& {Wright}}]{Vanderburg2015}
{Vanderburg}, A., {Johnson}, J.~A., {Rappaport}, S., {et~al.} 2015, \nat, 526,
  546, \dodoi{10.1038/nature15527}

\bibitem[{{Veras} {et~al.}(2022){Veras}, {Birader}, \& {Zaman}}]{Veras2022}
{Veras}, D., {Birader}, Y., \& {Zaman}, U. 2022, \mnras, 510, 3379,
  \dodoi{10.1093/mnras/stab3490}

\bibitem[{{Veras} {et~al.}(2017){Veras}, {Carter}, {Leinhardt}, \&
  {G{\"a}nsicke}}]{Veras2017}
{Veras}, D., {Carter}, P.~J., {Leinhardt}, Z.~M., \& {G{\"a}nsicke}, B.~T.
  2017, \mnras, 465, 1008, \dodoi{10.1093/mnras/stw2748}

\bibitem[{{Veras} \& {{\'C}uk}(2025)}]{Veras2025}
{Veras}, D., \& {{\'C}uk}, M. 2025, \mnras, \dodoi{10.1093/mnras/staf1120}

\bibitem[{{Veras} \& {Heng}(2020)}]{Veras&Hend2020}
{Veras}, D., \& {Heng}, K. 2020, \mnras, 496, 2292,
  \dodoi{10.1093/mnras/staa1632}

\bibitem[{{Veras} {et~al.}(2023){Veras}, {Ida}, {Grishin}, {Kenyon}, \&
  {Bromley}}]{Veras2023}
{Veras}, D., {Ida}, S., {Grishin}, E., {Kenyon}, S.~J., \& {Bromley}, B.~C.
  2023, \mnras, 524, 1, \dodoi{10.1093/mnras/stad1790}

\bibitem[{{Veras} {et~al.}(2014){Veras}, {Leinhardt}, {Bonsor}, \&
  {G{\"a}nsicke}}]{Veras2014}
{Veras}, D., {Leinhardt}, Z.~M., {Bonsor}, A., \& {G{\"a}nsicke}, B.~T. 2014,
  \mnras, 445, 2244, \dodoi{10.1093/mnras/stu1871}

\bibitem[{{Veras} {et~al.}(2020){Veras}, {McDonald}, \& {Makarov}}]{Veras2020}
{Veras}, D., {McDonald}, C.~H., \& {Makarov}, V.~V. 2020, \mnras, 492, 5291,
  \dodoi{10.1093/mnras/staa243}

\bibitem[{{Veras} {et~al.}(2024){Veras}, {Mustill}, \& {Bonsor}}]{Veras2024}
{Veras}, D., {Mustill}, A.~J., \& {Bonsor}, A. 2024, Reviews in Mineralogy and
  Geochemistry, 90, 141, \dodoi{10.2138/rmg.2024.90.05}

\bibitem[{{Verbunt} \& {Rappaport}(1988)}]{Verbunt1988}
{Verbunt}, F., \& {Rappaport}, S. 1988, \apj, 332, 193, \dodoi{10.1086/166645}

\bibitem[{{Vergely} {et~al.}(2022){Vergely}, {Lallement}, \&
  {Cox}}]{Vergely2022}
{Vergely}, J.~L., {Lallement}, R., \& {Cox}, N.~L.~J. 2022, \aap, 664, A174,
  \dodoi{10.1051/0004-6361/202243319}

\bibitem[{Virtanen {et~al.}(2020)Virtanen, Gommers, Oliphant, Haberland, Reddy,
  Cournapeau, Burovski, Peterson, Weckesser, Bright, {van der Walt}, Brett,
  Wilson, Millman, Mayorov, Nelson, Jones, Kern, Larson, Carey, Polat, Feng,
  Moore, {VanderPlas}, Laxalde, Perktold, Cimrman, Henriksen, Quintero, Harris,
  Archibald, Ribeiro, Pedregosa, {van Mulbregt}, \& {SciPy 1.0
  Contributors}}]{2020SciPy-NMeth}
Virtanen, P., Gommers, R., Oliphant, T.~E., {et~al.} 2020, Nature Methods, 17,
  261, \dodoi{10.1038/s41592-019-0686-2}

\bibitem[{{Wang} \& {Chen}(2019)}]{Wang2019}
{Wang}, S., \& {Chen}, X. 2019, \apj, 877, 116,
  \dodoi{10.3847/1538-4357/ab1c61}

\bibitem[{{Welsh} {et~al.}(2010){Welsh}, {Lallement}, {Vergely}, \&
  {Raimond}}]{Welsh2010}
{Welsh}, B.~Y., {Lallement}, R., {Vergely}, J.~L., \& {Raimond}, S. 2010, \aap,
  510, A54, \dodoi{10.1051/0004-6361/200913202}

\bibitem[{{Werner} {et~al.}(2021){Werner}, {Gorjian}, {Morales}, {Livingston},
  {Kennedy}, {Akeson}, {Beichman}, {Ciardi}, {Furlan}, {Lowrance}, {Mamajek},
  {Plavchan}, {Stark}, \& {Wyatt}}]{Werner2021}
{Werner}, M.~W., {Gorjian}, V., {Morales}, F.~Y., {et~al.} 2021, \apjs, 254,
  11, \dodoi{10.3847/1538-4365/abea20}

\bibitem[{{W}es {M}c{K}inney(2010)}]{mckinney-proc-scipy-2010}
{W}es {M}c{K}inney. 2010, in {P}roceedings of the 9th {P}ython in {S}cience
  {C}onference, ed. {S}t\'efan van~der {W}alt \& {J}arrod {M}illman, 56 -- 61,
  \dodoi{10.25080/Majora-92bf1922-00a}

\bibitem[{{Wilson} {et~al.}(2019){Wilson}, {Farihi}, {G{\"a}nsicke}, \&
  {Swan}}]{Wilson2019}
{Wilson}, T.~G., {Farihi}, J., {G{\"a}nsicke}, B.~T., \& {Swan}, A. 2019,
  \mnras, 487, 133, \dodoi{10.1093/mnras/stz1050}

\bibitem[{{Xu} {et~al.}(2019{\natexlab{a}}){Xu}, {Dufour}, {Klein}, {Melis},
  {Monson}, {Zuckerman}, {Young}, \& {Jura}}]{Xu2019b}
{Xu}, S., {Dufour}, P., {Klein}, B., {et~al.} 2019{\natexlab{a}}, \aj, 158,
  242, \dodoi{10.3847/1538-3881/ab4cee}

\bibitem[{{Xu} {et~al.}(2016){Xu}, {Jura}, {Dufour}, \& {Zuckerman}}]{Xu2016}
{Xu}, S., {Jura}, M., {Dufour}, P., \& {Zuckerman}, B. 2016, \apjl, 816, L22,
  \dodoi{10.3847/2041-8205/816/2/L22}

\bibitem[{{Xu} {et~al.}(2014){Xu}, {Jura}, {Koester}, {Klein}, \&
  {Zuckerman}}]{Xu2014}
{Xu}, S., {Jura}, M., {Koester}, D., {Klein}, B., \& {Zuckerman}, B. 2014,
  \apj, 783, 79, \dodoi{10.1088/0004-637X/783/2/79}

\bibitem[{{Xu} {et~al.}(2018){Xu}, {Rappaport}, {van Lieshout}, {Vanderburg},
  {Gary}, {Hallakoun}, {Ivanov}, {Wyatt}, {DeVore}, {Bayliss}, {Bento},
  {Bieryla}, {Cameron}, {Cann}, {Croll}, {Collins}, {Dalba}, {Debes}, {Doyle},
  {Dufour}, {Ely}, {Espinoza}, {Joner}, {Jura}, {Kaye}, {McClain}, {Muirhead},
  {Palle}, {Panka}, {Provencal}, {Randall}, {Rodriguez}, {Scarborough},
  {Sefako}, {Shporer}, {Strickland}, {Zhou}, \& {Zuckerman}}]{Xu2018a}
{Xu}, S., {Rappaport}, S., {van Lieshout}, R., {et~al.} 2018, \mnras, 474,
  4795, \dodoi{10.1093/mnras/stx3023}

\bibitem[{{Xu} {et~al.}(2019{\natexlab{b}}){Xu}, {Hallakoun}, {Gary}, {Dalba},
  {Debes}, {Dufour}, {Fortin-Archambault}, {Fukui}, {Jura}, {Klein},
  {Kusakabe}, {Muirhead}, {Narita}, {Steele}, {Su}, {Vanderburg}, {Watanabe},
  {Zhan}, \& {Zuckerman}}]{Xu2019a}
{Xu}, S., {Hallakoun}, N., {Gary}, B., {et~al.} 2019{\natexlab{b}}, \aj, 157,
  255, \dodoi{10.3847/1538-3881/ab1b36}

\bibitem[{{Zhou} {et~al.}(2016){Zhou}, {Kedziora-Chudczer}, {Bailey},
  {Marshall}, {Bayliss}, {Stockdale}, {Nelson}, {Tan}, {Rodriguez}, {Tinney},
  {Dragomir}, {Colon}, {Shporer}, {Bento}, {Sefako}, {Horne}, \&
  {Cochran}}]{Zhou2016}
{Zhou}, G., {Kedziora-Chudczer}, L., {Bailey}, J., {et~al.} 2016, \mnras, 463,
  4422, \dodoi{10.1093/mnras/stw2286}

\bibitem[{{Zuckerman}(2001)}]{Zuckerman2001}
{Zuckerman}, B. 2001, \araa, 39, 549, \dodoi{10.1146/annurev.astro.39.1.549}

\bibitem[{{Zuckerman} {et~al.}(2003){Zuckerman}, {Koester}, {Reid}, \&
  {H{\"u}nsch}}]{Zuckerman2003}
{Zuckerman}, B., {Koester}, D., {Reid}, I.~N., \& {H{\"u}nsch}, M. 2003, \apj,
  596, 477, \dodoi{10.1086/377492}

\bibitem[{{Zuckerman} {et~al.}(2010){Zuckerman}, {Melis}, {Klein}, {Koester},
  \& {Jura}}]{Zuckerman2010}
{Zuckerman}, B., {Melis}, C., {Klein}, B., {Koester}, D., \& {Jura}, M. 2010,
  \apj, 722, 725, \dodoi{10.1088/0004-637X/722/1/725}

\end{thebibliography}
\bibliographystyle{aasjournal}

\end{document}